# Hierarchical sparse Bayesian learning: theory and application for inferring structural damage from incomplete modal data

Yong Huang[1,2] and James L. Beck[1]

[1]Division of Engineering and Applied Science, California Institute of Technology, CA, USA

[2] Key Lab of Structural Dynamic Behavior and Control of the Ministry of Education, Harbin Institute of Technology, Harbin, China

**Abstract:**

Structural damage due to excessive loading or environmental degradation typically occurs in localized areas in the absence of collapse. This prior information about the spatial sparseness of structural damage is exploited here by a hierarchical sparse Bayesian learning framework with the goal of reducing the source of ill-conditioning in the stiffness loss inversion problem for damage detection. Sparse Bayesian learning methodologies automatically prune away irrelevant or inactive features from a set of potential candidates, and so they are effective probabilistic tools for producing sparse explanatory subsets. We have previously proposed such an approach to establish the probability of localized stiffness reductions that serve as a proxy for damage by using noisy incomplete modal data from before and after possible damage. The core idea centers on a specific hierarchical Bayesian model that promotes spatial sparseness in the inferred stiffness reductions in a way that is consistent with the Bayesian Ockham razor. In this paper, we improve the theory of our previously proposed sparse Bayesian learning approach by eliminating an approximation and, more importantly, incorporating a constraint on stiffness increases. Our approach has many appealing features that are summarized at the end of the paper. We validate the approach by applying it to the Phase II simulated and experimental benchmark studies sponsored by the IASC-ASCE Task Group on Structural Health Monitoring. The results show that it can reliably detect, locate and assess damage by inferring substructure stiffness losses from the identified modal parameters. The occurrence of missed and false damage alerts is effectively suppressed.

## 1. Introduction

The development of automated sensor-based structural health monitoring (SHM) systems for accurately detecting, locating and assessing damage from severe loading events (such as earthquake, hurricanes or explosions) or progressive structural deterioration (such as fatigue) has received much attention over the last several decades [e.g. 1-5]. Numerous vibration-based global SHM techniques have been developed that provide powerful tools for estimating both damage location and damage severity [1-2, 6-9], and which utilize the dependence of the identified structural modal parameters, such as natural frequencies and mode shapes, on physical properties of structures, i.e., stiffness, mass and damping. In practice, most SHM systems record structural vibration data at low-amplitude excitation, e.g., ambient vibration from wind and ground motion, and so a linear dynamic model with classical normal modes is adequate for damage detection purposes. The typical strategy involves comparing structural models identified from sets of measured modal data (natural frequencies and mode shapes) from a structure before and after possible damage [1-2, 6-8]. For applications in structural health monitoring, there are some inherent difficulties in this model-based approach.

One of the main difficulties is that it is impossible to exactly model the full behavior of a structure by using the limited knowledge available. For damage inference problems, there are always modeling uncertainties involved and hence the need to account for uncertainty in model updating arises. These uncertainties may arise from many sources: simplifying approximations to develop the structural models; limited number of sensors installed on a structure; sensor noise; thermally-induced daily variations in structural stiffness, etc. In this paper we tackle the problem of model updating and damage assessment from a Bayesian perspective and compute a Bayesian posterior belief (probability density function) for the structural model parameters [6, 14-19]. Rather than considering only a point estimate as in the conventional deterministic methods, Bayesian methods can be used to update both the relative plausibility of each value of the model parameters and each model within a set of candidates, enabling the consideration of all plausible models and the construction of robust predictions of future structural behavior [10].

Probability in the Bayesian sense is interpreted as the degree of plausibility or belief of a statement on the basis of the specific conditioning information [10-13], where the statement may refer to the structural parameters but also to the structural model itself. It quantifies uncertainty due to our incomplete information and does not need the postulate of "inherent randomness". The Bayesian probabilistic framework has been shown to be effective for structure damage detection and assessment [6, 14-19].

Another fundamental difficulty for damage detection lies in the fact that the mathematical nature of the underdetermined structural inverse problem based on the constraints imposed by the available data requires a tradeoff between the resolution of the damage locations and the reliability of the probabilistically-inferred damaged state. In this case, defining a proper threshold to determine whether the damage features shift from their healthy state will be very important in order to alleviate false positive (false alarm) and false negative (missed alarm) detections, though it is very challenging to compute a reliable threshold value in a rigorous manner for issuing a timely damage alarm [20,21]. In this paper, we aim to extend the applicability of Bayesian methods to produce reliable damage alarms even for high-dimensional model parameter spaces (higher-resolution damage localization), by the exploration of recent developments in sparse Bayesian learning [22,23] and Bayesian compressive sensing/sampling [24-26]. The new perspective for the stiffness loss inverse problem exploits the prior knowledge that structural damage typically occurs at a limited number of locations in a structure in the absence of its collapse. A specific hierarchical Bayesian model and the corresponding evidence maximization strategy for Bayesian inference is utilized to promote spatial sparseness in the inferred stiffness reductions in a way that is consistent with the Bayesian Ockham razor [10, 27].

The concept of system modal parameters, distinct from the corresponding modal parameters of the structural model, is employed. This is a useful strategy that avoids the matching of experimental and structural model modes in implementations [2,6,14,16-19]. Note that mode matching is required for most existing methods but it is nontrivial in practice, especially when incomplete mode shape measurements are available.

We have proposed previously to use a sparse Bayesian perspective to tackle the stiffness loss inverse problem [28,29]. In this paper, we present an improved version of our sparse Bayesian learning method that eliminates an approximation in our original theoretical formulation and incorporates an additional constraint to reflect the fact that the stiffness parameters cannot have values in a damaged state that are greater than those of the calibration state, which enhances its accuracy for damage detection and assessment. We also apply our method to real data, not only the synthetic data that we used previously [28,29].

The remainder of the paper is organized as follows. In Section 2, we present our hierarchical Bayesian model for the stiffness inversion problem. The hierarchical Bayesian model is then used in an evidence maximization procedure in Section 3, yielding a Bayesian inference framework that lends itself to efficient computation of the posterior probability density function for the structural model parameters. In Section 4, two variants of the algorithm are proposed. Applications of our methods to simulated and real structural data are presented in Section 5, and conclusions and future work are discussed in Section 6.

## 2. Hierarchical Bayesian model class

Suppose that $N_s$ sets of measured vibration time histories are available from a structure and $N_m$ modes of vibration have been identified for each set of time histories so that we have a vector of identified natural frequencies: $\hat{\boldsymbol{\omega}}^2 = [\hat{\omega}_{1,1}^2, \ldots, \hat{\omega}_{1,N_m}^2, \hat{\omega}_{2,1}^2, \ldots, \hat{\omega}_{N_s,N_m}^2]^T \in \mathbb{R}^{N_s N_m \times 1}$ and identified mode shapes: $\hat{\boldsymbol{\psi}} = [\hat{\boldsymbol{\psi}}_{1,1}^T, \ldots, \hat{\boldsymbol{\psi}}_{1,N_m}^T, \hat{\boldsymbol{\psi}}_{2,1}^T, \ldots, \hat{\boldsymbol{\psi}}_{N_s,N_m}^T]^T \in \mathbb{R}^{N_s N_m N_o \times 1}$, where $\hat{\boldsymbol{\psi}}_{r,i} \in \mathbb{R}^{N_o}$ gives the identified components of the system mode shape of the $i^{th}$ mode ($i = 1, \ldots, N_m$) at the $N_o$ measured DOF (degrees of freedom) from the $r^{th}$ data segment ($r = 1, \ldots, N_s$). These modal parameters are assumed to be directly estimated from dynamic data using an appropriate modal identification procedure, such as MODE-ID [16, 30], which does not use a structural model and identifies the maximum a posteriori (MAP) values of the natural frequencies $\hat{\boldsymbol{\omega}}^2$, the mode shape components $\hat{\boldsymbol{\psi}}$ at the observed DOFs and the equivalent viscous damping ratios for each significant mode.

The $N_o$ measured DOF in a structure are usually a smaller subset of the $N_d$ DOF of an appropriate structural model. It has been shown to be advantageous [2,6,14,16-19] to introduce the system natural frequencies $\boldsymbol{\omega}^2 = [\omega_1^2, \ldots, \omega_{N_m}^2]^T \in \mathbb{R}^{N_m \times 1}$ and system mode shapes $\boldsymbol{\phi} = [\boldsymbol{\phi}_1^T, \ldots, \boldsymbol{\phi}_{N_m}^T]^T \in \mathbb{R}^{N_d N_m \times 1}$ to represent the actual underlying modal parameters of the linear dynamics of the structural system at all DOFs corresponding to those of the structural model. In addition, following [19], we use the eigenvalue equations of the structural model as "soft" constraints in a prior PDF (probability density function). Based on a defined *stochastic model class* $\mathcal{M}$ (see [10]), one can use the available modal data $\hat{\boldsymbol{\omega}}^2$ and $\hat{\boldsymbol{\psi}}$ to update the system modal parameters $\boldsymbol{\omega}^2$ and $\boldsymbol{\phi}$ together with structural model parameters $\boldsymbol{\theta}$, without using mode matching.

*2.1 Structural model class*

As noted earlier, we use a linear dynamic model with classical normal modes (so a damping matrix need not be explicitly modeled) because we use low-amplitude vibration data recorded by the structural monitoring system just before and after possible damage occurrence. As part of the definition of the stochastic model class $\mathcal{M}$, a set of linear structural models is first taken with $N_d$ DOFs where each model has the same known mass matrix $\mathbf{M} \in \mathbb{R}^{N_d \times N_d}$ (which is assumed to be determined with sufficient accuracy from structural drawings) and an uncertain stiffness matrix $\mathbf{K}$ parameterized as a linear combination of $(N_\theta + 1)$ substructure stiffness matrices $\mathbf{K}_j, j = 0, 1, \ldots N_\theta$:

$$\mathbf{K}(\boldsymbol{\theta}) = \mathbf{K}_0 + \sum_{j=1}^{N_\theta} \theta_j \mathbf{K}_j \tag{1}$$

where $\mathbf{K}_0$ is any part of the global stiffness matrix that is considered known and that corresponds to parts of the structure that are believed to be not vulnerable to damage, and $\mathbf{K}_j \in \mathbb{R}^{N_d \times N_d}, j = 1, \ldots, N_\theta$, is the substructure stiffness matrix representing the nominal contribution of the $j^{th}$ substructure to the global stiffness matrix $\mathbf{K}$ from a structural model (e.g., specified by a finite element model of a structure). The corresponding stiffness scaling parameter $\theta_j, j = 1, \ldots, N_\theta$, is a factor that allows modification of the nominal $j^{th}$ substructure contribution and any reduction of $\theta_j, j = 1, \ldots N_\theta$, is assumed to correspond to damage in the $j^{th}$ substructure. The parameter vector $\boldsymbol{\theta} = [\theta_1, \ldots, \theta_n] \in \mathbb{R}^{N_\theta}$, therefore specifies the structural model in $\mathcal{M}$. Since the onset of damage is typically in a small number of locations in a structure in the absence of structural collapse, $\Delta \boldsymbol{\theta} = \boldsymbol{\theta} - \hat{\boldsymbol{\theta}}_u$ can be considered as a sparse vector with relatively few non-zero components, where $\boldsymbol{\theta}$ and $\boldsymbol{\theta}_u$ are the stiffness scaling parameters for the current state (possibly damaged) and the calibration (undamaged) state, and $\hat{\boldsymbol{\theta}}_u$ is the MAP estimate of $\boldsymbol{\theta}_u$ from applying a Bayesian updating method without utilizing sparseness to the modal data from the undamaged structure.

*2.2 Prior distribution for system modal parameters and structural model parameters*

Continuing with the definition of the stochastic model class $\mathcal{M}$, a joint prior PDF is specified for the system modal parameters and structural model parameters to reflect the relative plausibility of their values in the absence of any measurement data. We do not assume that the system natural frequencies $\boldsymbol{\omega}^2 = [\omega_1^2, \dots, \omega_{N_m}^2]^T \in \mathbb{R}^{N_m \times 1}$ and system mode shapes $\boldsymbol{\phi} = [\boldsymbol{\phi}_1^T, \dots, \boldsymbol{\phi}_{N_m}^T]^T \in \mathbb{R}^{N_m N_d \times 1}$ satisfy the eigenvalue problem corresponding to any structural model specified by $\boldsymbol{\theta}$ because there will always be modeling errors $\mathbf{e} = [\boldsymbol{e}_1^T, \dots, \boldsymbol{e}_{N_m}^T] \in \mathbb{R}^{N_m N_d}$, where $(\mathbf{K}(\boldsymbol{\theta}) - \omega_i^2 \mathbf{M})\boldsymbol{\phi}_i = \mathbf{e}_i$, $i = 1, \dots, N_m$, for the $i^{th}$ system mode and the structural model specified by $\boldsymbol{\theta}$. The uncertain eigenvalue equation error $\mathbf{e}$ therefore provides a bridge between the behaviors of the real system and a deterministic structural model.

Using the Principle of Maximum Information Entropy [12,13], we model the $\mathbf{e}_i$, $i = 1, \dots, N_m$, as zero-mean independent and identically distributed Gaussian vectors with covariance matrix $\beta^{-1}\mathbf{I}_d = \text{diag}(\beta^{-1}, \dots, \beta^{-1})$, since this gives a joint probability model that corresponds to the largest uncertainty for $\mathbf{e}$ (i.e., it maximizes Shannon's information entropy subject to the first two moment constraints: $\mathbf{E}[(\mathbf{e}_i)_k] = 0, \mathbf{E}[(\mathbf{e}_i)_k^2] = \beta^{-1}$, $k = 1, \dots, N_d$, $i = 1, \dots, N_m$). Based on the specified Gaussian probability model for $\mathbf{e}_i$, $i = 1, \dots, N_m$, the following prior PDF conditional on the precision hyper-parameter $\beta$ is created:

$$p(\boldsymbol{\omega}^2, \boldsymbol{\phi}, \boldsymbol{\theta}|\beta) = c_0 (2\pi/\beta)^{-N_m N_d/2} \exp\left\{-\frac{\beta}{2}\sum_{i=1}^{N_m}\left\|(\mathbf{K}(\boldsymbol{\theta}) - \omega_i^2 \mathbf{M})\boldsymbol{\phi}_i\right\|^2\right\} \tag{2}$$

where $c_0$ is a normalizing constant and $\|\cdot\|$ denotes the Euclidean vector norm, so $\|\mathbf{x}\|^2 = \mathbf{x}^T\mathbf{x}$. Note that the equation-error precision parameter $\beta$ in (2) allows for the explicit control of how closely the system and model modal parameters agree. However, it is difficult to choose an appropriate value a priori for $\beta$ and this motivates the introduction later of a hierarchical Bayesian prior (see Equation 7), where an optimal value of $\beta$ is learned from the data. Notice that as $\beta \to \infty$, the system modal parameters become tightly clustered around the modal parameters corresponding to the structural model specified by $\boldsymbol{\theta}$, which are given as $(\mathbf{K}(\boldsymbol{\theta}) - \omega^2\mathbf{M})\boldsymbol{\phi} = \mathbf{0}$. Note also that if $\boldsymbol{\theta}$ is specified then these model modal parameters are always the most plausible values a priori of the system modal parameters.

The exponent in (2) is a quadratic in $\boldsymbol{\theta}$ and so (2) can be analytically integrated with respect to $\boldsymbol{\theta}$ to get the marginal prior PDF for the system modal parameters $[\boldsymbol{\omega}^2, \boldsymbol{\phi}]$ as:

$$p(\boldsymbol{\omega}^2, \boldsymbol{\phi}|\beta) = c_0 (2\pi/\beta)^{(N_\theta - N_d N_m)/2} |\mathbf{H}^T \mathbf{H}|^{-1/2} \exp\left\{-\frac{\beta}{2}(\mathbf{b}^T \mathbf{b} - \mathbf{b}^T \mathbf{H}(\mathbf{H}^T \mathbf{H})^{-1} \mathbf{H}^T \mathbf{b})\right\} \tag{3}$$

where the parameter vector $\mathbf{b}$ and matrix $\mathbf{H}$ are defined by:

$$\mathbf{b} = \begin{bmatrix} (\omega_1^2 \mathbf{M} - \mathbf{K}_0)\boldsymbol{\phi}_1 \\ \vdots \\ (\omega_{N_m}^2 \mathbf{M} - \mathbf{K}_0)\boldsymbol{\phi}_{N_m} \end{bmatrix}_{N_m N_d \times 1} \tag{4}$$

$$\mathbf{H} = \begin{bmatrix} \mathbf{K}_1 \boldsymbol{\phi}_1 & \cdots & \mathbf{K}_{N_\theta} \boldsymbol{\phi}_1 \\ \vdots & \ddots & \vdots \\ \mathbf{K}_1 \boldsymbol{\phi}_{N_m} & \cdots & \mathbf{K}_{N_\theta} \boldsymbol{\phi}_{N_m} \end{bmatrix}_{N_m N_d \times N_\theta} \tag{5}$$

We can deduce the prior PDF for $\boldsymbol{\theta}$ conditional on system modal parameters $[\boldsymbol{\omega}^2, \boldsymbol{\phi}]$ from (2) and (3) as:

$$p(\boldsymbol{\theta}|\boldsymbol{\omega}^2, \boldsymbol{\phi}, \beta) = p(\boldsymbol{\omega}^2, \boldsymbol{\phi}, \boldsymbol{\theta}|\beta)/p(\boldsymbol{\omega}^2, \boldsymbol{\phi}|\beta) = \mathcal{N}(\boldsymbol{\theta}|(\mathbf{H}^T \mathbf{H})^{-1}\boldsymbol{\Psi}^T \mathbf{b}, (\beta \mathbf{H}^T \mathbf{H})^{-1})$$

$$= (2\pi/\beta)^{-N_\theta/2} |\mathbf{H}^T \mathbf{H}|^{1/2} \exp\left\{-\frac{\beta}{2}(\boldsymbol{\theta} - (\mathbf{H}^T \mathbf{H})^{-1}\mathbf{H}^T \mathbf{b})^T \mathbf{H}^T \mathbf{H}(\boldsymbol{\theta} - (\mathbf{H}^T \mathbf{H})^{-1}\mathbf{H}^T \mathbf{b})\right\} \tag{6}$$

We model the uncertainty in $\beta$ by the following widely used Gamma hyperprior:

$$p(\beta|a_0, b_0) = \text{Gam}(\beta|a_0, b_0) = \frac{b_0^{-a_0}}{\Gamma(a_0)} \beta^{a_0 - 1} \exp(-b_0^{-1}\beta) \tag{7}$$

where $a_0$ and $b_0$ are shape and scale parameters, respectively. The choice of $a_0$ is discussed later in Section 3.2. To penalize values of $b_0$ that are too large, we assign an exponential hyper-prior for scale parameter $b_0$ as the last stage of the hierarchical model:

$$p(b_0|\kappa) = \text{Exp}(b_0|\kappa) = \kappa \exp(-\kappa b_0) \tag{8}$$

*2.3 Likelihood function for structural model parameters*

During the calibration stage, we assume that the model is a globally identifiable [31] in $\boldsymbol{\theta}_u$, the structural stiffness scaling parameter vector for the calibration state, and that the Bayesian updating of $\boldsymbol{\theta}_u$ uses a large amount of time-domain vibration data, so that a unique MAP estimate $\widehat{\boldsymbol{\theta}}_u$ is obtained. The connection between the structural stiffness scaling parameter vector $\boldsymbol{\theta}$ at the current monitoring state and that at the calibration state is given as:

$$\boldsymbol{\theta} = \widehat{\boldsymbol{\theta}}_u + \Delta\boldsymbol{\theta} \tag{9}$$

where $\Delta\boldsymbol{\theta}$ is the potential change in the parameter vector from damage and it is modeled as a zero-mean

Gaussian vector, based on the Principle of Maximum Information Entropy[12,13]. As we argued previously, the stiffness parameter changes $\Delta\boldsymbol{\theta} = \boldsymbol{\theta} - \widehat{\boldsymbol{\theta}}_u$ should be a sparse vector (most of its components zero). We can encode a preference for sparseness in $\Delta\boldsymbol{\theta}$ by employing a variance $\alpha_j$ associated independently with every component $\theta_j$ and choosing the MAP value $\widehat{\boldsymbol{\theta}}_u$ from the calibration state as pseudo-data for $\boldsymbol{\theta}$ to define a likelihood function as:

$$p(\widehat{\boldsymbol{\theta}}_u|\boldsymbol{\theta},\boldsymbol{\alpha}) = \mathcal{N}(\widehat{\boldsymbol{\theta}}_u|\boldsymbol{\theta},\mathbf{A}) = \prod_{j=1}^{N_\theta} \mathcal{N}(\hat{\theta}_{u,j}|\theta_j,\alpha_j) \tag{10}$$

where $\mathbf{A} = \text{diag}(\alpha_1,\dots,\alpha_{N_\theta})$. The idea here is to learn each $\alpha_j$ from the modal data and if $\alpha_j \to 0$, then $\theta_j \to \hat{\theta}_{u,j}$, which is interpreted as the $j^{th}$ substructure being undamaged. Equation (10) is the only connection between the calibration state and the monitoring stage. The choice of likelihood function in (10) is motivated by the closely-related sparse Bayesian learning framework which is known to provide an effective tool for pruning large numbers of irrelevant or redundant features in a linear regression model that are not supported by the data [22,23]. Although the conventional strategy in sparse Bayesian learning is to use an automatic relevance determination (ARD) Gaussian prior PDF [27] to model sparseness, here we incorporate the ARD concept in the likelihood function, along with the prior on $\boldsymbol{\theta}$ in (6). This choice still leads to a sparse representation of the parameter change vector $\Delta\boldsymbol{\theta}$ during the optimization of the hyper-parameter vector $\boldsymbol{\alpha}$ using an evidence maximization procedure [22,23].

*2.4 Likelihood functions for system modal parameters*

To complete the definition of the stochastic model class $\mathcal{M}$, the likelihood function for the system modal parameters $\boldsymbol{\omega}^2$ and $\boldsymbol{\phi}$ is assigned to express the probability of getting modal data $\widehat{\boldsymbol{\omega}}^2$ and $\widehat{\boldsymbol{\psi}}$ when $\boldsymbol{\omega}^2$ and $\boldsymbol{\phi}$ are given. Using the Principle of Maximum Information Entropy again, the combined prediction errors and measurement errors for the system modal parameters $\boldsymbol{\omega}^2$ and $\boldsymbol{\phi}$ are modeled independently as zero-mean Gaussian variables with unknown variances, so:

$$p(\widehat{\boldsymbol{\omega}}^2,\widehat{\boldsymbol{\psi}}|\boldsymbol{\omega}^2,\boldsymbol{\phi},\boldsymbol{\theta}) = p(\widehat{\boldsymbol{\omega}}^2|\boldsymbol{\omega}^2)p(\widehat{\boldsymbol{\psi}}|\boldsymbol{\phi})$$

$$= \mathcal{N}(\widehat{\boldsymbol{\omega}}^2|\mathbf{L}\boldsymbol{\omega}^2,\mathbf{E})\,\mathcal{N}(\widehat{\boldsymbol{\psi}}|\boldsymbol{\Gamma}\boldsymbol{\phi},\mathbf{C}) \tag{11}$$

where the selection matrix $\boldsymbol{\Gamma} \in \mathbb{R}^{N_o N_m N_s \times N_m N_d}$ with "1s" and "0s" picks the observed DOF of the whole "measured" mode shape data set $\widehat{\boldsymbol{\psi}}$ from the system mode shapes $\boldsymbol{\phi}$; $\mathbf{L} = [\mathbf{I}_{N_m},\dots,\mathbf{I}_{N_m}]^T \in \mathbb{R}^{N_0 N_m \times N_m}$ is the transformation matrix between the vector of $N_s$ sets of $N_m$ identified natural frequencies $\widehat{\boldsymbol{\omega}}^2$ and the $N_m$ system natural frequencies $\boldsymbol{\omega}^2$; $\mathbf{E} = \text{block diag}(\mathbf{E}_1, \ldots, \mathbf{E}_{N_s})$ is a block diagonal covariance matrix with the diagonal block $\mathbf{E}_r = \text{diag}(\rho_1^{-1}, \ldots, \rho_{N_m}^{-1})$, $r = 1, \ldots, N_s$; and $\mathbf{C} = \eta^{-1}\mathbf{I}_{N_oN_sN_m}$; $\mathbf{I}_{N_m}$ and $\mathbf{I}_{N_oN_sN_m}$ denote the identity matrices of corresponding size; and $\boldsymbol{\rho} = [\rho_1, \ldots, \rho_{N_m}]^T$ and $\eta$ are prescribed precision parameters for the predictions of the identified natural frequencies $\widehat{\boldsymbol{\omega}}^2$ and mode shapes $\widehat{\boldsymbol{\psi}}$ from the system modal parameters, respectively. In a hierarchical manner, exponential priors are placed on the parameters $\rho_i$ and $\eta$:

$$p(\rho_i|\tau_i) = \text{Exp}(\rho_i|\tau_i) = \tau_i \exp(-\tau_i \rho_i), \quad i = 1, \ldots, N_m$$

$$p(\eta|\nu) = \text{Exp}(\eta|\nu) = \nu \exp(-\nu\eta) \tag{12}$$

which are the maximum entropy priors with support $[0, \infty)$ for given mean values $\tau_i^{-1}$ and $\nu^{-1}$ of $\rho_i$ and $\eta$, respectively. Then the prior PDF for the parameter vector $\boldsymbol{\rho}$ is given by:

$$p(\boldsymbol{\rho}|\boldsymbol{\tau}) = \prod_{i=1}^{N_m} p(\rho_i|\tau_i) = \prod_{i=1}^{N_m} \tau_i \cdot \exp\left(-\sum_{i=1}^{N_m} \tau_i \rho_i\right) \tag{13}$$

where $\boldsymbol{\tau} = [\tau_1, \ldots, \tau_{N_m}]^T$.

*2.5 Joint posterior PDF for hierarchical Bayesian model*

By combining all stages of the hierarchical Bayesian model, the joint posterior PDF of all the uncertain-valued parameters conditional on the observed quantities is finally achieved by using Bayes' theorem:

$$p(\boldsymbol{\omega}^2, \boldsymbol{\rho}, \boldsymbol{\tau}, \boldsymbol{\phi}, \eta, \nu, \boldsymbol{\theta}, \boldsymbol{\alpha}, \beta, a_0, b_0, \kappa | \widehat{\boldsymbol{\omega}}^2, \widehat{\boldsymbol{\psi}}, \widehat{\boldsymbol{\theta}}_u)$$

$$\propto p(\widehat{\boldsymbol{\omega}}^2|\boldsymbol{\omega}^2, \boldsymbol{\rho}) p(\widehat{\boldsymbol{\psi}}|\boldsymbol{\phi}, \eta) p(\widehat{\boldsymbol{\theta}}_u|\boldsymbol{\theta}, \boldsymbol{\alpha}) p(\boldsymbol{\omega}^2, \boldsymbol{\phi}, \boldsymbol{\theta}|\beta) p(\boldsymbol{\rho}|\boldsymbol{\tau}) p(\eta|\nu) p(\boldsymbol{\alpha}) p(\boldsymbol{\tau}) p(\nu)$$

$$\cdot p(\beta|a_0, b_0) p(b_0|\kappa) p(a_0) p(\kappa) \tag{14}$$

where the PDFs $p(\widehat{\boldsymbol{\omega}}^2|\boldsymbol{\omega}^2, \boldsymbol{\rho}) p(\widehat{\boldsymbol{\psi}}|\boldsymbol{\phi}, \eta)$, $p(\widehat{\boldsymbol{\theta}}_u|\boldsymbol{\theta}, \boldsymbol{\alpha})$, $p(\boldsymbol{\omega}^2, \boldsymbol{\phi}, \boldsymbol{\theta}|\beta)$, $p(\boldsymbol{\rho}|\boldsymbol{\tau})$, $p(\eta|\nu)$, $p(\beta|a_0, b_0)$ and $p(b_0|\kappa)$ are defined in (11), (10), (2), (13), (12), (7) and (8), respectively. The product of PDFs $p(\boldsymbol{\alpha})p(\boldsymbol{\tau})p(\nu)p(a_0)p(\kappa)$ is a constant because each of the priors is chosen as a broad uniform PDF, and the product is henceforth absorbed into the proportionality constant in (14).

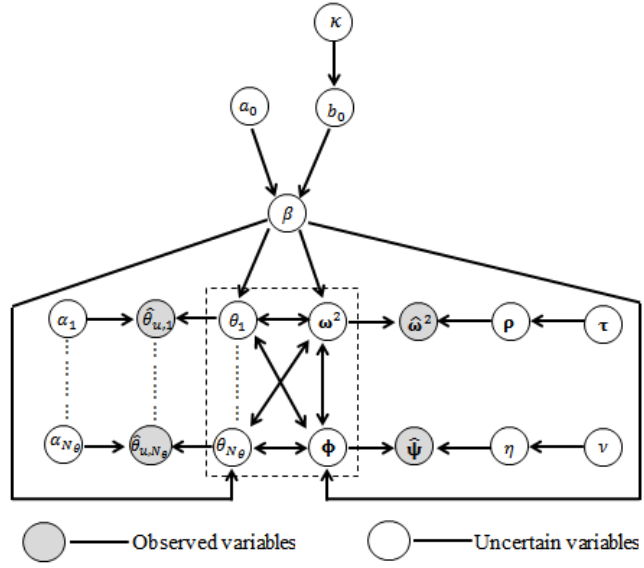

**Fig 1.** Acyclic graph representing the hierarchical Bayesian model.

The acyclic graph of the hierarchical Bayesian model is shown in Figure 1, where each arrow denotes the conditional dependencies used in the joint probability model. Compared with non-hierarchical models where $\beta, \boldsymbol{\rho}$ and $\eta$ are selected a priori [19], the hierarchical Bayesian models allow all sources of uncertainty and correlation to be learned from the data, and hence they potentially produce more accurate parameter identification.

### 3. Bayesian inference

*3.1 Approximation of the full posterior PDF using Laplace's method*

The structural stiffness scaling parameter $\boldsymbol{\theta}$ is the key focus of the stiffness inversion problem. For convenience, we denote the other uncertain parameters by the vector $\boldsymbol{\delta} = [(\boldsymbol{\omega}^2)^T, \boldsymbol{\rho}^T, \boldsymbol{\tau}^T, \boldsymbol{\phi}^T, \eta^T, \nu^T, \boldsymbol{\alpha}^T, \beta, a_0, b_0, \kappa]^T$, then our Bayesian inference is based on the full posterior PDF over all uncertain parameters from Bayes' theorem:

$$p(\boldsymbol{\theta}, \boldsymbol{\delta} | \widehat{\boldsymbol{\omega}}^2, \widehat{\boldsymbol{\psi}}, \widehat{\boldsymbol{\theta}}_u) = \frac{p(\boldsymbol{\theta}, \boldsymbol{\delta}, \widehat{\boldsymbol{\omega}}^2, \widehat{\boldsymbol{\psi}}, \widehat{\boldsymbol{\theta}}_u)}{p(\widehat{\boldsymbol{\omega}}^2, \widehat{\boldsymbol{\psi}}, \widehat{\boldsymbol{\theta}}_u)} \tag{15}$$

The denominator in (15) is the marginalized probability of the data according to the hierarchical model:

$$p(\widehat{\boldsymbol{\omega}}^2, \widehat{\boldsymbol{\psi}}, \widehat{\boldsymbol{\theta}}_u) = \int p(\boldsymbol{\theta}, \boldsymbol{\delta}, \widehat{\boldsymbol{\omega}}^2, \widehat{\boldsymbol{\psi}}, \widehat{\boldsymbol{\theta}}_u) d\boldsymbol{\theta} d\boldsymbol{\delta} \tag{16}$$

and is nearly always an analytically intractable integral. Nevertheless, we can turn to a hierarchical Bayesian procedure combined with Laplace's asymptotic approximation [31] to find an effective approximation of the full posterior PDF in (15).

Using the probability product rule, we rewrite the full posterior in (15) as:

$$p(\boldsymbol{\theta}, \boldsymbol{\delta}|\widehat{\boldsymbol{\omega}}^2, \widehat{\boldsymbol{\psi}}, \widehat{\boldsymbol{\theta}}_u) = p(\boldsymbol{\theta}|\boldsymbol{\delta}, \widehat{\boldsymbol{\omega}}^2, \widehat{\boldsymbol{\psi}}, \widehat{\boldsymbol{\theta}}_u)p(\boldsymbol{\delta}|\widehat{\boldsymbol{\omega}}^2, \widehat{\boldsymbol{\psi}}, \widehat{\boldsymbol{\theta}}_u) \tag{17}$$

The first factor in (17) is the posterior PDF of the stiffness scaling parameter vector conditional on $\boldsymbol{\delta}$, which is computed from Bayes' Theorem by:

$$p(\boldsymbol{\theta}|\boldsymbol{\delta}, \widehat{\boldsymbol{\omega}}^2, \widehat{\boldsymbol{\psi}}, \widehat{\boldsymbol{\theta}}_u) = p(\boldsymbol{\theta}|\boldsymbol{\delta}, \widehat{\boldsymbol{\theta}}_u) = p(\widehat{\boldsymbol{\theta}}_u|\boldsymbol{\theta}, \boldsymbol{\delta})p(\boldsymbol{\theta}|\boldsymbol{\delta})/p(\widehat{\boldsymbol{\theta}}_u|\boldsymbol{\delta}) \tag{18}$$

where the first equality holds because $\boldsymbol{\theta}$ is independent of $\widehat{\boldsymbol{\omega}}^2$ and $\widehat{\boldsymbol{\psi}}$ when conditional on $\boldsymbol{\omega}^2$ and $\boldsymbol{\phi}$ in $\boldsymbol{\delta}$ (see Figure 1) and the second equality is due to Bayes' Theorem. As a consequence of combining a Gaussian prior (6) and a linear model within a Gaussian likelihood (10), the conditional posterior is also Gaussian:

$$p(\boldsymbol{\theta}|\boldsymbol{\delta}, \widehat{\boldsymbol{\omega}}^2, \widehat{\boldsymbol{\psi}}, \widehat{\boldsymbol{\theta}}_u) = \mathcal{N}(\boldsymbol{\theta}|\boldsymbol{\mu}, \boldsymbol{\Sigma}) \tag{19}$$

with mean and covariance matrix:

$$\boldsymbol{\mu} = \boldsymbol{\Sigma}(\beta \mathbf{H}^T \mathbf{b} + \mathbf{A}^{-1}\widehat{\boldsymbol{\theta}}_u) \tag{20a}$$

$$\boldsymbol{\Sigma} = (\beta \mathbf{H}^T \mathbf{H} + \mathbf{A}^{-1})^{-1} \tag{20b}$$

The normalizing constant for the posterior PDF in (18) is called the *pseudo-evidence* (or *pseudo-marginal likelihood*) *for the model class* $\mathcal{M}(\boldsymbol{\delta})$ given by pseudo-data $\widehat{\boldsymbol{\theta}}_u$ and it can be analytically evaluated:

$$p(\widehat{\boldsymbol{\theta}}_u|\boldsymbol{\delta}) = \int p(\widehat{\boldsymbol{\theta}}_u|\boldsymbol{\theta}, \boldsymbol{\delta})p(\boldsymbol{\theta}|\boldsymbol{\delta})d\boldsymbol{\theta} = \int p(\widehat{\boldsymbol{\theta}}_u|\boldsymbol{\theta}, \boldsymbol{\alpha})\, p(\boldsymbol{\theta}|\boldsymbol{\omega}^2, \boldsymbol{\phi}, \beta)d\boldsymbol{\theta} = \mathcal{N}(\widehat{\boldsymbol{\theta}}_u|(\mathbf{H}^T\mathbf{H})^{-1}\mathbf{H}^T\mathbf{b}, \mathbf{D}) \tag{21}$$

where $\mathbf{D} = \mathbf{A} + \beta^{-1}(\mathbf{H}^T\mathbf{H})^{-1}$ and we have utilized the hierarchical structure exhibited in Figure 1. This is a Gaussian distribution over the $N_\theta$-dimensional pre-damage parameter vector $\boldsymbol{\theta}_u$ evaluated at its MAP value $\widehat{\boldsymbol{\theta}}_u$, and is readily evaluated for arbitrary values of $\boldsymbol{\delta}$.

The second factor in (17) is given by Bayes' Theorem:

$$p(\boldsymbol{\delta}|\widehat{\boldsymbol{\omega}}^2, \widehat{\boldsymbol{\psi}}, \widehat{\boldsymbol{\theta}}_u) = \frac{p(\widehat{\boldsymbol{\omega}}^2, \widehat{\boldsymbol{\psi}}, \widehat{\boldsymbol{\theta}}_u|\boldsymbol{\delta})p(\boldsymbol{\delta})}{p(\widehat{\boldsymbol{\omega}}^2, \widehat{\boldsymbol{\psi}}, \widehat{\boldsymbol{\theta}}_u)} \tag{22}$$

Assuming that the problem is globally identifiable, meaning here that the posterior $p(\boldsymbol{\delta}|\widehat{\boldsymbol{\omega}}^2, \widehat{\boldsymbol{\psi}}, \widehat{\boldsymbol{\theta}}_u)$ in (22) has a unique maximum at $\widetilde{\boldsymbol{\delta}}$ (the MAP value of $\boldsymbol{\delta}$), we treat $\boldsymbol{\delta}$ in (17) as a 'nuisance' parameter vector and integrate it out using Laplace's asymptotic approximation [31]:

$$p(\boldsymbol{\theta}|\widehat{\boldsymbol{\omega}}^2, \widehat{\boldsymbol{\psi}}, \widehat{\boldsymbol{\theta}}_u) = \int p(\boldsymbol{\theta}|\boldsymbol{\delta}, \widehat{\boldsymbol{\omega}}^2, \widehat{\boldsymbol{\psi}}, \widehat{\boldsymbol{\theta}}_u) p(\boldsymbol{\delta}|\widehat{\boldsymbol{\omega}}^2, \widehat{\boldsymbol{\psi}}, \widehat{\boldsymbol{\theta}}_u) \, d\boldsymbol{\delta}$$

$$\approx p(\boldsymbol{\theta}|\widetilde{\boldsymbol{\delta}}, \widehat{\boldsymbol{\omega}}^2, \widehat{\boldsymbol{\psi}}, \widehat{\boldsymbol{\theta}}_u) \tag{23}$$

where $\widetilde{\boldsymbol{\delta}} = \arg\max p(\boldsymbol{\delta}|\widehat{\boldsymbol{\omega}}^2, \widehat{\boldsymbol{\psi}}, \widehat{\boldsymbol{\theta}}_u) = \arg\max[p(\widehat{\boldsymbol{\omega}}^2, \widehat{\boldsymbol{\psi}}, \widehat{\boldsymbol{\theta}}_u|\boldsymbol{\delta}) \cdot p(\boldsymbol{\delta})]$ because the denominator in (22) is independent of $\boldsymbol{\delta}$. In the next section, we will focus on the task of maximizing $p(\boldsymbol{\delta}|\widehat{\boldsymbol{\omega}}^2, \widehat{\boldsymbol{\psi}}, \widehat{\boldsymbol{\theta}}_u)$ for finding the MAP value $\widetilde{\boldsymbol{\delta}}$.

*3.2 MAP estimate of $\boldsymbol{\delta}$ from PDF $p(\boldsymbol{\delta}|\widehat{\boldsymbol{\omega}}^2, \widehat{\boldsymbol{\psi}}, \widehat{\boldsymbol{\theta}}_u)$*

Finding the MAP value $\widetilde{\boldsymbol{\delta}}$ is equivalent to maximizing with respect to $\boldsymbol{\delta}$:

$$J_1(\boldsymbol{\delta}) = p(\widehat{\boldsymbol{\omega}}^2, \widehat{\boldsymbol{\psi}}, \widehat{\boldsymbol{\theta}}_u|\boldsymbol{\delta}) p(\boldsymbol{\delta})$$

$$= p(\widehat{\boldsymbol{\omega}}^2|\boldsymbol{\omega}^2, \boldsymbol{\rho}) p(\widehat{\boldsymbol{\psi}}|\boldsymbol{\phi}, \eta) p(\widehat{\boldsymbol{\theta}}_u|\boldsymbol{\omega}^2, \boldsymbol{\phi}, \beta, \boldsymbol{\alpha}) p(\boldsymbol{\omega}^2, \boldsymbol{\phi}|\beta) p(\boldsymbol{\rho}|\boldsymbol{\tau}) p(\eta|\nu) p(\beta|a_0, b_0) p(b_0|\kappa) \tag{24}$$

where we have used the hierarchical structure exhibited in Figure 1.

Substituting equations (11), (21), (3), (13), (12), (7) and (8), we get:

$$J_1(\boldsymbol{\delta}) = c_0 \cdot \mathcal{N}(\widehat{\boldsymbol{\omega}}^2|\mathbf{L}\boldsymbol{\omega}^2, \mathbf{E}) \cdot \mathcal{N}(\widehat{\boldsymbol{\psi}}|\boldsymbol{\Gamma}\boldsymbol{\phi}, \mathbf{C}) \cdot \mathcal{N}(\widehat{\boldsymbol{\theta}}_u|(\mathbf{H}^T\mathbf{H})^{-1}\mathbf{H}^T\mathbf{b}, \mathbf{D})$$

$$\cdot \beta^{(N_d N_m - N_\theta)/2} \cdot |\mathbf{H}^T\mathbf{H}|^{-1/2} \cdot \nu \cdot \prod_{i=1}^{N_m} \tau_i \cdot \text{Gam}(\beta|a_0, b_0) \cdot \kappa$$

$$\cdot \exp\left[-\frac{\beta}{2}(\mathbf{b}^T\mathbf{b} - \mathbf{b}^T\mathbf{H}(\mathbf{H}^T\mathbf{H})^{-1}\mathbf{H}^T\mathbf{b}) - \sum_{i=1}^{N_m}(\tau_i \rho_i) - \nu\eta - \kappa b_0\right] \tag{25}$$

Instead of maximizing $J_1(\boldsymbol{\delta})$ in (25), we maximize equivalently its logarithm $\log J_1(\boldsymbol{\delta})$ with respect to $\boldsymbol{\delta}$. It is shown in the Appendix that explicit expressions can be obtained to be used for iterative maximization of $\log J_1(\boldsymbol{\delta})$ where we update all of the parameters in $\boldsymbol{\delta}$ successively and repeat until convergence.

By maximizing $\log J_1(\boldsymbol{\delta})$ with respect to the system mode shapes $\boldsymbol{\phi}$ and the associated hyper-parameters $\eta$ and $\nu$ successively, with all other parameters fixed at their MAP values, the MAP estimates $\widetilde{\boldsymbol{\phi}}, \widetilde{\eta}$ and $\widetilde{\nu}$ are derived as (see Appendix):

$$\widetilde{\boldsymbol{\phi}} = \left(\widetilde{\beta}\widetilde{\mathbf{F}}^T\widetilde{\mathbf{F}} + \widetilde{\beta}\,\text{diag}\left(\text{tr}(\widetilde{\boldsymbol{\Sigma}}\mathbf{T}_1), \ldots, \text{tr}(\widetilde{\boldsymbol{\Sigma}}\mathbf{T}_{N_m N_d})\right) + \boldsymbol{\Gamma}^T\widetilde{\mathbf{C}}^{-1}\boldsymbol{\Gamma}\right)^{-1}\left(\boldsymbol{\Gamma}^T\widetilde{\mathbf{C}}^{-1}\widehat{\boldsymbol{\psi}} - \widetilde{\beta}[\text{tr}(\widetilde{\boldsymbol{\Sigma}}\mathbf{U}_1), \ldots, \text{tr}(\widetilde{\boldsymbol{\Sigma}}\mathbf{U}_{N_m N_d})]^T\right) \tag{26}$$

$$\widetilde{\eta} = \frac{N_s N_m N_o}{2\widetilde{\nu} + \|\widehat{\boldsymbol{\psi}} - \boldsymbol{\Gamma}\widetilde{\boldsymbol{\phi}}\|^2} \tag{27}$$

$$\tilde{v} = 1/\tilde{\eta} \tag{28}$$

where $\widetilde{\boldsymbol{\Sigma}}$ is given by (20b) using $\tilde{\beta}$, $\widetilde{\boldsymbol{\alpha}}$ and $\widetilde{\boldsymbol{\phi}}$ in **H** in (5),

$$\widetilde{\mathbf{F}} = \begin{bmatrix} \mathbf{K}(\widetilde{\boldsymbol{\mu}}) - \widetilde{\omega}_1^2 \mathbf{M} & \cdots & \mathbf{0} \\ \vdots & \ddots & \vdots \\ \mathbf{0} & \cdots & \mathbf{K}(\widetilde{\boldsymbol{\mu}}) - \widetilde{\omega}_1^2 \mathbf{M} \end{bmatrix}_{N_m N_d \times N_m N_d} \tag{29}$$

where $\widetilde{\boldsymbol{\mu}}$ is given by (20a) using $\tilde{\beta}$, $\widetilde{\boldsymbol{\alpha}}$ and also $\widetilde{\boldsymbol{\phi}}, \widetilde{\boldsymbol{\omega}}^2$ in **b** and **H** in (4) and (5), and for $q = 1, \ldots, N_m N_d$:

$$\mathbf{T}_q = \mathbf{\Pi}_q^T \mathbf{\Pi}_q$$

$$\mathbf{U}_q = \mathbf{\Pi}_q^T \cdot \left( \sum_{s \neq q}^{N_m N_d} \tilde{\phi}_s \mathbf{\Pi}_s \right)$$

$$\mathbf{\Pi}_q = \frac{\partial \mathbf{H}}{\partial \phi_q} \tag{30}$$

We substitute (28) into (27) and find that the optimal estimate of $\tilde{\eta}$ has components:

$$\tilde{\eta} = \frac{N_s N_m N_o - 2}{\|\widehat{\boldsymbol{\psi}} - \boldsymbol{\Gamma} \widetilde{\boldsymbol{\phi}}\|^2} \tag{31}$$

Similarly, the MAP estimates of the system natural frequencies $\boldsymbol{\omega}^2$ and their associated hyper-parameters $\boldsymbol{\rho}$ and $\boldsymbol{\tau}$ are also derived as (see Appendix):

$$\widetilde{\boldsymbol{\omega}}^2 = \left[ \mathbf{L}^T \widetilde{\mathbf{E}}^{-1} \mathbf{L} + \tilde{\beta} \widetilde{\mathbf{G}}^T \widetilde{\mathbf{G}} \right]^{-1} \left( \mathbf{L}^T \widetilde{\mathbf{E}}^{-1} \widehat{\boldsymbol{\omega}}^2 + \tilde{\beta} \widetilde{\mathbf{G}}^T \tilde{\mathbf{c}} \right) \tag{32}$$

and for each $i = 1, \ldots, N_m$:

$$\tilde{\rho}_i = \frac{N_s}{2\tilde{\tau}_i + \sum_{r=1}^{N_o} (\widehat{\omega}_{r,i}^2 - \widetilde{\omega}_i^2)^2} \tag{33}$$

$$\tilde{\tau}_i = 1/\tilde{\rho}_i \tag{34}$$

where:

$$\widetilde{\mathbf{G}} = \begin{bmatrix} \mathbf{M}\widetilde{\boldsymbol{\phi}}_1 & \cdots & 0 \\ \vdots & \ddots & \vdots \\ 0 & \cdots & \mathbf{M}\widetilde{\boldsymbol{\phi}}_{N_m} \end{bmatrix}_{N_d N_m \times N_m} \tag{35}$$

$$\tilde{\mathbf{c}} = \begin{bmatrix} \mathbf{K}(\widetilde{\boldsymbol{\mu}})\widetilde{\boldsymbol{\phi}}_1 \\ \vdots \\ \mathbf{K}(\widetilde{\boldsymbol{\mu}})\widetilde{\boldsymbol{\phi}}_{N_m} \end{bmatrix}_{N_d N_m \times 1} \tag{36}$$

We solve for $\tilde{\rho}_i$ using (34) in (33) for each $i = 1, \ldots, N_m$:

$$\tilde{\rho}_i = \frac{N_s - 2}{\sum_{r=1}^{N_o} (\widehat{\omega}_{r,i}^2 - \widetilde{\omega}_i^2)^2} \tag{37}$$

Similarly, the MAP estimates of the hyper-parameters $\tilde{\alpha}_j, j = 1, \ldots, N_\theta$, are derived as (see Appendix):

$$\tilde{\alpha}_j = \left(\widetilde{\boldsymbol{\Sigma}}\right)_{jj} + \left(\widehat{\boldsymbol{\theta}}_u - \widetilde{\boldsymbol{\mu}}\right)_j^2 \tag{38}$$

Finally, the MAP values of $\beta, a_0, b_0$ and $\kappa$ are obtained by taking the derivatives of (25) with respect to them (see Appendix):

$$\tilde{\beta} = \frac{N_d N_m - \sum_{j=1}^{N_\theta}\left(1 - \tilde{\alpha}_j^{-1}\Sigma_{jj}\right) + 2(\tilde{a}_0 - 1)}{\|\widetilde{\mathbf{H}}\boldsymbol{\mu} - \widetilde{\mathbf{b}}\|^2 + 2\tilde{b}_0^{-1}} \tag{39}$$

$$\log(\tilde{a}_0 \tilde{b}_0 - \tilde{\kappa}) - \log(\tilde{b}_0) - \psi(\tilde{a}_0) = 0 \tag{40}$$

$$\tilde{b}_0 = \frac{2\tilde{\beta}}{\tilde{a}_0 + \sqrt{\tilde{a}_0^2 + 4\tilde{\kappa}\tilde{\beta}}} \tag{41}$$

$$\tilde{\kappa} = 1/\tilde{b}_0 \tag{42}$$

where $\psi(a_0)$ is the Digamma function. There is no analytical solution for the optimal estimate $\tilde{a}_0$ in (40), so instead, we set $\tilde{a}_0 = 1$ in the algorithm later to produce the exponential distribution for $\beta$, which is the maximum entropy prior with the constraint of $\mathbf{E}(\beta|b_0) = b_0$.

The final MAP estimate $\widetilde{\boldsymbol{\delta}}$ is calculated by iterating through the sequence of equations (26), (31) (32), (37), (38), (39) and (41-42).

*3.3 Discussion of calculation of MAP estimates of hyper-parameters*

For maximization of the pseudo-evidence $p(\widehat{\boldsymbol{\theta}}_u|\boldsymbol{\delta})$ with respect to $\boldsymbol{\delta}$, there is an information-theoretical interpretation of the trade-off between data fitting and model complexity [10,26] that can be demonstrated as follows, where we use Bayes' Theorem and the hierarchical model structure in Figure 1:

$$\log[p(\widehat{\boldsymbol{\theta}}_u|\boldsymbol{\delta})] = \int \log[p(\widehat{\boldsymbol{\theta}}_u|\boldsymbol{\omega}^2,\boldsymbol{\phi},\beta,\boldsymbol{\alpha})] p(\boldsymbol{\theta}|\widehat{\boldsymbol{\theta}}_u,\boldsymbol{\omega}^2,\boldsymbol{\phi},\beta,\boldsymbol{\alpha})\, d\boldsymbol{\theta}$$

$$= \int \log[p(\boldsymbol{\theta}|\boldsymbol{\omega}^2,\boldsymbol{\phi},\beta)\, p(\widehat{\boldsymbol{\theta}}_u|\boldsymbol{\theta},\boldsymbol{\alpha})/p(\boldsymbol{\theta}|\widehat{\boldsymbol{\theta}}_u,\boldsymbol{\omega}^2,\boldsymbol{\phi},\beta,\boldsymbol{\alpha})] p(\boldsymbol{\theta}|\widehat{\boldsymbol{\theta}}_u,\boldsymbol{\omega}^2,\boldsymbol{\phi},\beta,\boldsymbol{\alpha})\, d\boldsymbol{\theta}$$

$$= \int \log[p(\widehat{\boldsymbol{\theta}}_u|\boldsymbol{\theta},\boldsymbol{\alpha})] p(\boldsymbol{\theta}|\widehat{\boldsymbol{\theta}}_u,\boldsymbol{\omega}^2,\boldsymbol{\phi},\beta,\boldsymbol{\alpha})d\boldsymbol{\theta} - \int \log[p(\boldsymbol{\theta}|\widehat{\boldsymbol{\theta}}_u,\boldsymbol{\omega}^2,\boldsymbol{\phi},\beta,\boldsymbol{\alpha})/p(\boldsymbol{\theta}|\boldsymbol{\omega}^2,\boldsymbol{\phi},\beta)] p(\boldsymbol{\theta}|\widehat{\boldsymbol{\theta}}_u,\boldsymbol{\omega}^2,\boldsymbol{\phi},\beta,\boldsymbol{\alpha})d\boldsymbol{\theta} \tag{43}$$

Equation (43) shows that the log pseudo-evidence, which is to be maximized, is the difference between the posterior mean of the log likelihood function for $\boldsymbol{\theta}$ (the first term) and the relative entropy (or Kullback-Leibler information) of the posterior with respect to the prior distribution for $\boldsymbol{\theta}$ (the second term). The first term quantifies the ability of the hierarchical model to match the MAP vector $\widehat{\boldsymbol{\theta}}_u$ which is obtained from the calibration state; it is maximized if all $\theta_j$ are tightly clustered around $\hat{\theta}_{u,j}$ in the posterior for $\boldsymbol{\theta}$ (i.e. all $\alpha_j \to 0$). The second term reflects the amount of information extracted from the pseudo-data $\widehat{\boldsymbol{\theta}}_u$; it is minimized if $\boldsymbol{\theta}$ is independent of $\widehat{\boldsymbol{\theta}}_u$, conditional on $(\boldsymbol{\omega}^2, \boldsymbol{\phi}, \beta)$ (i.e. all $\alpha_j \to \infty$). This term penalizes models that have fewer parameter components $\theta_j$ differing from those in $\widehat{\boldsymbol{\theta}}_u$, and therefore forces the model updating to extract less information from $\widehat{\boldsymbol{\theta}}_u$ and relatively more from the system modal parameters $\boldsymbol{\omega}^2$ and $\boldsymbol{\phi}$, and so from the "measured" modal data $\widehat{\boldsymbol{\omega}}^2$ and $\widehat{\boldsymbol{\psi}}$, as implied by the observation of the hierarchical model structure exhibited in Figure 1. However, over-extraction of information from the system modal parameters $\boldsymbol{\omega}^2$ and $\boldsymbol{\phi}$ will produce a structural model vector $\boldsymbol{\theta}$ with too large of a difference from $\widehat{\boldsymbol{\theta}}_u$ that is overly sensitive to the details of the information in the specified system modal parameters $\boldsymbol{\omega}^2$ and $\boldsymbol{\phi}$, and so the "measured" modal parameters $\widehat{\boldsymbol{\omega}}^2$ and $\widehat{\boldsymbol{\psi}}$. In other words, the sensor noise and other environmental effects may not be "smoothed out" so they may have a large effect on the damage detection performance.

To summarise, (43) demonstrates that the evidence maximization procedure over hyper-parameter vector $\boldsymbol{\delta}$ automatically involves a trade-off between the average data-fit of the model class defined by $\boldsymbol{\delta}$ and the information it extracts from the data. Roughly speaking, the first term in (43) is larger if $\widehat{\boldsymbol{\theta}}_u$ dominates $(\boldsymbol{\omega}^2, \boldsymbol{\phi}, \beta)$ in providing information about $\boldsymbol{\theta}$ whereas the penalty from the second term is smaller if $(\boldsymbol{\omega}^2, \boldsymbol{\phi}, \beta)$ dominates $\widehat{\boldsymbol{\theta}}_u$ similarly. Maximizing the evidence to give the MAP value of $\boldsymbol{\delta}$ therefore forces a compromise between these two competing effects. This is the Principle of Model Parsimony or the Bayesian Ockham razor [10, 27] at work. The Bayesian procedure is effectively implementing Ockham's Razor by assigning lower probabilities to a structural model whose parameter vector $\boldsymbol{\theta}$ has too large or too small differences from $\widehat{\boldsymbol{\theta}}_u$ obtained at the calibration state (too few or too many $\alpha_j \to 0$), therefore suppressing the occurrence of false and missed damage detection alarms.

*3.4 Evaluation of damage probability*

Compared with the deterministic sparse inversion algorithms that are based on least-squares with regularization using the $l_1$ norm, such as Basis Pursuit [32-33] and Orthogonal Matching Pursuit [34], the advantage of the sparse Bayesian method proposed here is that it not only provides the most probable estimate of the stiffness parameter vector to specify the sparse representation but also the corresponding posterior uncertainty. The latter allows for a probabilistic damage measure to be calculated [2,6].

Given the identified modal parameters $[\widehat{\boldsymbol{\omega}}_d^2, \widehat{\boldsymbol{\Psi}}_d]$ and $[\widehat{\boldsymbol{\omega}}_u^2, \widehat{\boldsymbol{\Psi}}_u]$ from the current monitoring stage where damage is possible and from the undamaged calibration stage, respectively, the posterior PDF of the stiffness scaling parameters $\boldsymbol{\theta}$ can be used to quantify the probability that the $j^{th}$ substructure stiffness scaling parameter has been reduced by more than a specified fraction $f$ of its value in the initial calibration stage. To proceed, we denote the stiffness scaling parameter of the $j^{th}$ substructure for the current possibly damaged state and initial undamaged state as $\theta_{d,j}$ and $\theta_{u,j}$, respectively, and assume that they are conditionally independent based on the modal data. By using a Gaussian asymptotic approximation [31, 35] for the integrals involved, we get [2],

$$P_j^{dam}(f) = P(\theta_{d,j} < (1-f)\theta_{u,j} | \widehat{\boldsymbol{\omega}}_u^2, \widehat{\boldsymbol{\Psi}}_u, \widehat{\boldsymbol{\omega}}_d^2, \widehat{\boldsymbol{\Psi}}_d)$$

$$= \int P(\theta_{d,j} < (1-f)\theta_{u,j} | \theta_{u,j}, \widehat{\boldsymbol{\omega}}_d^2, \widehat{\boldsymbol{\Psi}}_d) p(\theta_{u,j} | \widehat{\boldsymbol{\omega}}_u^2, \widehat{\boldsymbol{\Psi}}_u) d\theta_{u,j}$$

$$\approx \Phi\left[\frac{(1-f)\mu_{u,j} - \mu_{d,j}}{\sqrt{(1-f)^2 \sigma_{d,j}^2 + \sigma_{u,j}^2}}\right] \tag{44}$$

where $\Phi(\cdot)$ is the standard Gaussian cumulative distribution function; $\mu_{d,j}$ and $\mu_{u,j}$ denote the posterior means of the stiffness scaling parameters of the $j^{th}$ substructure for the possibly damaged and undamaged structure, respectively, given by (20a); and $\sigma_{d,j}$ and $\sigma_{u,j}$ are the corresponding posterior standard deviations of the stiffness scaling parameters of the $j^{th}$ substructure, which are the square root of the diagonal elements of the posterior covariance matrix $\boldsymbol{\Sigma}$ given by (20b). Equation (20a) and (20b) are applied for $\theta_{d,j}$ and $\theta_{u,j}$ using the modal data $[\widehat{\boldsymbol{\omega}}_d^2, \widehat{\boldsymbol{\Psi}}_d]$ and $[\widehat{\boldsymbol{\omega}}_u^2, \widehat{\boldsymbol{\Psi}}_u]$, respectively. For this calculation, we do not impose a constraint on stiffness increases from the calibration stage (i.e., $\theta_{d,j} > \theta_{u,j}$ is allowed here, although $\mu_{u,j} \leq \mu_{d,j}$ always from Algorithm 2 presented later).

## 4. Proposed algorithms

The above formulation leads to the following algorithms for probabilistic inference of structural stiffness loss. There are two variants of the algorithm. For the calibration stage, model sparseness is not expected and hence Algorithm 1 is used without optimization of the hyper-parameter vector $\boldsymbol{\alpha}$; instead, we fix all components $\alpha_j$ with some large values. For the monitoring stage, Algorithm 2 is used where all hyper-parameters in $\boldsymbol{\delta}$ are estimated to ensure sparse stiffness loss inference.

*4.1 Algorithm 1: Bayesian inference for the structural model in the calibration stage*

---

**Algorithm 1**

---

INPUTS:

▷ Identified modal data $\widehat{\boldsymbol{\omega}}^2 = \widehat{\boldsymbol{\omega}}_u^2$ and $\widehat{\boldsymbol{\psi}} = \widehat{\boldsymbol{\psi}}_u$ from the calibration stage

▷ Initial values $\bar{\eta}, \bar{\boldsymbol{\rho}}$ and $\bar{\beta}$, and chosen value of $b_0$

▷ A chosen nominal value $\boldsymbol{\theta}_0$ of $\boldsymbol{\theta}_u$ (e.g. from a finite-element model) and associated prior covariance matrix $\boldsymbol{\Sigma}_0 = \sigma_0^2 \mathbf{I}_{N_\theta}$ with large variance $\sigma_0^2$.

ALGORITHM:

1. Initialize $\boldsymbol{\theta} = \boldsymbol{\theta}_0, \boldsymbol{\Sigma} = \boldsymbol{\Sigma}_0$ and $\widetilde{\boldsymbol{\omega}}^2 = \sum_{r=1}^{N_s} \widehat{\boldsymbol{\omega}}_r^2 / N_s$

2. Initialize $\eta, \boldsymbol{\rho}$ and $\beta$ as the initial values $\bar{\eta}, \bar{\boldsymbol{\rho}}$ and $\bar{\beta}$, respectively

3. Fix all components in $\widetilde{\boldsymbol{\alpha}}$ with large values (e.g. $\tilde{\alpha}_j = 10^9$)

4. **While** convergence criterion on $\boldsymbol{\theta}$ is not met

5. Update MAP $\widetilde{\boldsymbol{\phi}}$ using (26), and then update $\tilde{\eta}$ using (31)

6. Update MAP $\widetilde{\boldsymbol{\omega}}^2$ using (32), and then update $\widetilde{\boldsymbol{\rho}}$ using (37)

7. Update MAP $\tilde{\beta}$ using (39), and then update $\tilde{b}_0$ and $\tilde{\kappa}$ using (41) and (42), respectively

8. Calculate the conditional posterior mean $\widehat{\boldsymbol{\theta}}_u = \boldsymbol{\mu}$ and covariance matrix $\boldsymbol{\Sigma}_u = \boldsymbol{\Sigma}$ for $\boldsymbol{\theta} = \boldsymbol{\theta}_u$ using (20a) and (20b)

9. **End while** (convergence criterion has been satisfied)

---

OUTPUTS:

▷ MAP estimates of all uncertain parameters in $\boldsymbol{\delta}$

▷ Posterior mean $\widehat{\boldsymbol{\theta}}_u$ and covariance matrix $\boldsymbol{\Sigma}_u$ of the stiffness scaling parameters $\boldsymbol{\theta}_u$ at the calibration stage

*4.2 Algorithm 2: Sparse Bayesian learning for inference of structural stiffness reductions in the monitoring stage*

When real data from a structure is used, the modeling uncertainty in the structural stiffness may be too large to have the stiffness reduction reliably detected by most of existing methods. To regularize a possibly severe ill-posed inverse problem, Ching and Beck [16][18] imposed a priori constraint in their Expectation–Maximization algorithm that the values of the stiffness parameters for a damaged state cannot exceed those for the calibration state. However, we have not found an analytically tractable solution for sparse Bayesian learning when imposing this additional prior information in the formulation. Therefore, we present an alternative strategy: $\alpha_j's$ for those substructures having $\tilde{\theta}_j > \hat{\theta}_{u,j}$ for the current iteration will be set to zero directly (the first two iterations will be discarded to make sure that the parameters in $\boldsymbol{\delta}$ are not directly influenced by the initial setting). Then the prior information that any stiffness increase from the calibration model is unacceptable is effectively utilized in our sparse Bayesian learning procedure for inference of structural stiffness reductions in the monitoring stage, which is summarized in Algorithm 2.

**Algorithm 2**

INPUTS:

▷ Identified modal data $\widehat{\boldsymbol{\omega}}^2 = \widehat{\boldsymbol{\omega}}_d^2$ and $\widehat{\boldsymbol{\psi}} = \widehat{\boldsymbol{\psi}}_d$ from the monitoring stage

▷ Initial values $\overline{\boldsymbol{\eta}}, \overline{\boldsymbol{\rho}}$ and $\bar{\beta}$, and chosen value of $b_0$

▷ The MAP estimate $\widehat{\boldsymbol{\theta}}_u$ and the associated posterior covariance matrix $\boldsymbol{\Sigma}_u$ from the calibration stage

▷ The MAP estimates $\widetilde{\boldsymbol{\omega}}_u^2$ and $\widetilde{\boldsymbol{\phi}}_u$ of the system natural frequencies and mode shapes from the calibration stage

ALGORITHM:

1. Initialize $\tilde{\alpha}_j = N_\theta^2$, $j = 1, \ldots, N_\theta$

2. Initialize $\widetilde{\boldsymbol{\omega}}^2$ and $\widetilde{\boldsymbol{\phi}}$ as their MAP estimates from the calibration stage

3. Initialize $\eta, \boldsymbol{\rho}$ and $\beta$ as the initial values $\bar{\eta}, \overline{\boldsymbol{\rho}}$ and $\bar{\beta}$, respectively

4. **While** convergence criterion on $\boldsymbol{\alpha}$ is not met

5. Set $\tilde{\alpha}_j = 0$ if $\alpha_j$ ($j = 1, \ldots, N_\theta$) becomes smaller than $\alpha_{min}$ (chosen as $10^{-9}$ in the example later)

6. Set $\tilde{\alpha}_j = 0$ if $\tilde{\theta}_j > \hat{\theta}_{u,j}$ ($j = 1, \ldots, N_\theta$) when the iteration number $l > 2$

7. Calculate the conditional posterior MAP value $\widetilde{\boldsymbol{\theta}}_d = \boldsymbol{\mu}$ and covariance matrix $\boldsymbol{\Sigma}_d = \boldsymbol{\Sigma}$ using (20a) and (20b)

8. Update MAP $\widetilde{\boldsymbol{\phi}}$ using (26), and then update $\tilde{\eta}$ using (31)

9. Update MAP $\widetilde{\boldsymbol{\omega}}^2$ using (32), and then update $\widetilde{\boldsymbol{\rho}}$ using (37)

10. Update MAP $\tilde{\beta}$ using (39), and then update $\tilde{b}_0$ and $\tilde{\kappa}$ using (41) and (42), respectively

11. Update $\widetilde{\boldsymbol{\alpha}}$ using (38)

12. **End while** (convergence criterion has been satisfied)

**OUTPUTS:**

▷ MAP estimates of all uncertain parameters in $\boldsymbol{\delta}$

▷ Posterior mean $\widetilde{\boldsymbol{\theta}}_d$ and covariance matrix $\boldsymbol{\Sigma}_d$ of the stiffness scaling parameters $\boldsymbol{\theta}_d$ at the monitoring stage

*Remark 4.1:* Algorithm 2 is performed by iterating between two groups of parameters $\boldsymbol{\theta}$ and $\boldsymbol{\delta}$, which can be regarded as a generalized version of the algorithm for sparse Bayesian learning in [22]. Our procedure starts by considering all substructures as possibly damaged (the reason for the choice $\tilde{\alpha}_j = N_\theta^2$ at the first iteration, $j = 1, \ldots, N_\theta$) and then causes the "inactive" components $\theta_j$, which have $\tilde{\alpha}_j < 10^{-9}$, to be exactly equal to $\hat{\theta}_{u,j}$ from the calibration stage when optimizing over the hyper-parameters $\alpha_j$, so that finally there are only a few "active" $\theta_j$'s that are changed from their calibration values $\hat{\theta}_{u,j}$; their corresponding substructures are considered to be damaged.

*Remark 4.2:* Compared with the proposed algorithm in [29], the posterior uncertainty in the stiffness scaling parameters $\boldsymbol{\theta}$ is explicitly incorporated for the learning of the MAP estimates of all other uncertain parameters $\boldsymbol{\delta} = \left[(\boldsymbol{\omega}^2)^T, \boldsymbol{\rho}^T, \boldsymbol{\tau}^T, \boldsymbol{\phi}^T, \boldsymbol{\eta}^T, \boldsymbol{\nu}^T, \boldsymbol{\alpha}^T, \beta, a_0, b_0, \kappa\right]^T$ using an evidence maximization procedure in Algorithm 2. It is expected that this will produce more sparse stiffness reduction models.

*Remark 4.3:* In Algorithm 2, the solution of parameter $\boldsymbol{\delta}$ is obtained by solving the following problem:

$$\max \ p(\boldsymbol{\delta}|\widehat{\boldsymbol{\omega}}^2, \widehat{\boldsymbol{\psi}}, \widehat{\boldsymbol{\theta}}_u), \ \text{subject to} \ \mu_j \leq \hat{\theta}_{u,j}(j = 1, \ldots, N_\theta)$$

where posterior mean $\mu$ is a function of uncertain $\delta$. Even though it is a "greedy" optimization by setting $\tilde{\alpha}_j = 0$ directly if any $\tilde{\theta}_j > \hat{\theta}_{u,j}$ ($j = 1, \ldots, N_\theta$) for each iteration (once an $\alpha_j = 0$, it stays at zero), the strategy will suppress those local maxima which correspond to models having some substructures with stiffness increases from the calibration state and hence significantly increase the probability of finding the global maximum with respect to $\boldsymbol{\delta}$. In addition, under the condition that the modeling errors are zero-mean, there is a possible compensation between the modeling error-induced stiffness increases and decreases from the calibration structural model, and so by suppressing stiffness increases, the strategy will help to reduce the occurrence of inaccurate stiffness reductions and also induce a higher degree of sparseness in the identified stiffness reduction results.

*Remark 4.4:* In order to enforce sparseness more strongly, a hierarchical form of the Laplace prior $p(\boldsymbol{\alpha}|\lambda)$ is utilized in [29] and it works well for stiffness loss identification with the simulated Phase II benchmark data studied in [29]. For real-data structural identification with large modeling errors, however, we found that it is unreliable to use this hyperprior to adjust the strength of the model sparseness. It can produce over-sparse results with too few stiffness reductions, thereby missing some damage. This is because the information extraction of the structural model parameter $\boldsymbol{\theta}$ from system modal parameters $\tilde{\boldsymbol{\omega}}^2$ and $\tilde{\boldsymbol{\phi}}$, and so the "measured" modal parameters $\widehat{\boldsymbol{\omega}}^2$ and $\widehat{\boldsymbol{\psi}}$, is insensitive to the sparseness level of the stiffness reduction when the modeling error is large; hence when finding the MAP values of $\boldsymbol{\alpha}$ by maximizing $p(\boldsymbol{\alpha}|\widehat{\boldsymbol{\omega}}^2, \widehat{\boldsymbol{\psi}}, \widehat{\boldsymbol{\theta}}_u, \lambda) \propto p(\widehat{\boldsymbol{\omega}}^2, \widehat{\boldsymbol{\psi}}, \widehat{\boldsymbol{\theta}}_u|\lambda)p(\boldsymbol{\alpha}|\lambda)$, the Laplace hyperprior $p(\boldsymbol{\alpha}|\lambda)$ will dominate over the evidence $p(\widehat{\boldsymbol{\omega}}^2, \widehat{\boldsymbol{\psi}}, \widehat{\boldsymbol{\theta}}_u|\boldsymbol{\alpha})$. This is the reason why a uniform hyper-prior over $\boldsymbol{\alpha}$ ($\lambda = 0$), rather than hierarchical Laplace prior as in [29], is assigned in this paper.

*4.4 Implementation details for Algorithms 1-2*

(1) *Hyper-parameters initialization.* Using (27), (33) and (39) along with some approximations, the hyper-parameters $\eta$, $\rho_i$ and $\beta$ are initialized in the iterative calculation of the MAP values as:

$$\bar{\eta} = (N_s N_m N_o - 2)/\|\widehat{\boldsymbol{\Psi}}\|^2, \tag{45}$$

$$\bar{\rho}_i = (N_o - 2)/\sum_{r=1}^{N_o} \widehat{\omega}_{r,i}^4, i = 1, \ldots, N_m \tag{46}$$

$$\bar{\beta} = \frac{1}{2} b_0 \big(N_d N_m + 2(a_0 - 1)\big) \tag{47}$$

(2) *Convergence criterion.* The iterations of Algorithm 2 are terminated when the change in all $\log \alpha_j$'s between the $(l-1)^{th}$ iteration and the $l^{th}$ iteration are sufficiently small (e.g. smaller than 0.005). For Algorithm 1, the convergence criterion is satisfied when the change in all components of $\boldsymbol{\mu}$ is sufficiently small (e.g. smaller than 0.001).

(3) *Number of data segments utilized.* It is seen from (37) that the number of data-segments for modal parameter identification should be at least three ($N_s \geq 3$) for tractable estimations of the hyper-parameters $\widetilde{\boldsymbol{\rho}}$.

## 5. Illustrative applications

The applicability of the proposed methodology to identify the substructure stiffness reductions is illustrated with the IASC-ASCE Phase II SHM benchmark problems, first using the simulated data [36], and second using the experimental data [37]. The first application serves to verify the proposed methodology while the second one illustrates its application to real data from a damaged structure. The benchmark structure is a four-story, two-bay by two-bay steel braced-frame and a diagram for the analytical model is depicted in Fig. 2 along with its dimensions, in which the *x*-direction is the strong direction of the columns. A detailed description of the the benchmark problem including detailed nominal properties of the structural elements in the analytical model can be found in [17].

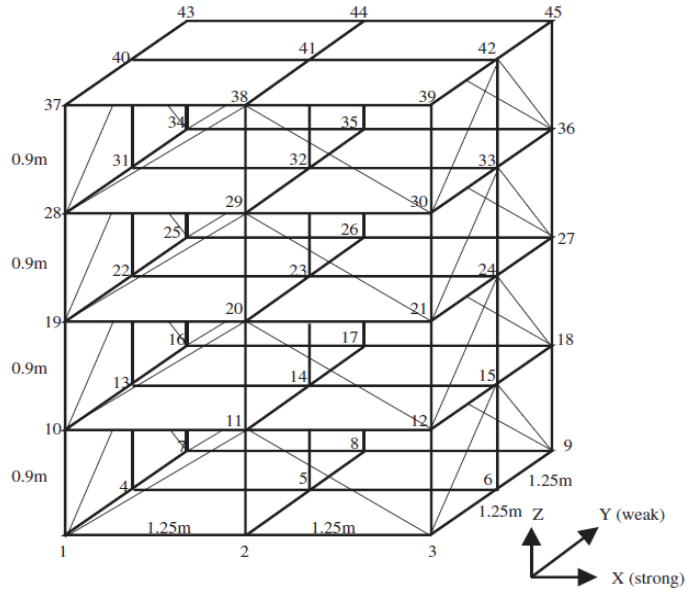

**Fig. 2.** The diagram of the benchmark structure [17,18].

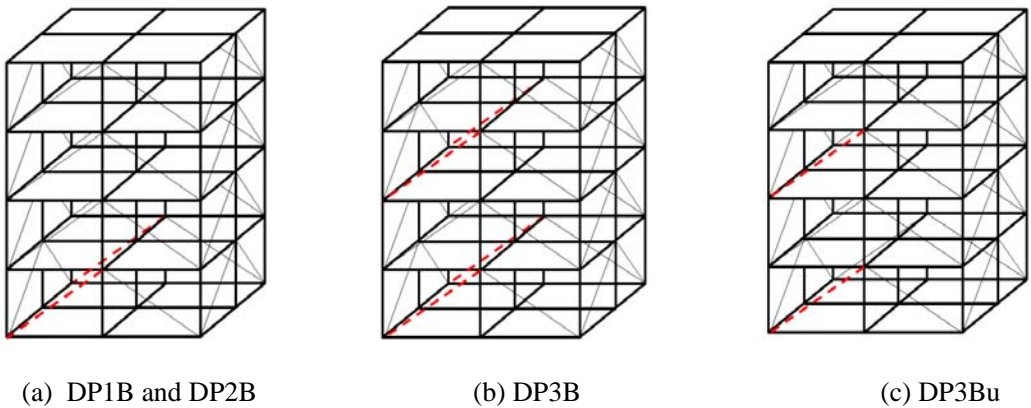

(a) DP1B and DP2B   (b) DP3B   (c) DP3Bu

**Fig. 3.** Damage patterns for simulated brace-damage cases (the dashed lines indicate the corresponding damaged locations) [17,18].

*5.1. Simulated Phase II benchmark problem*

Simulated Phase II benchmark problem is defined in Bernal et al. [36] and consists of both brace and joint damage cases, but only the brace cases are studied here. Four cases of brace damage are simulated by reducing the Young's moduli of certain braces in the structural model. Using br X-Y to denote the brace joining nodes X and Y in Figure 2: 1) DP1B: 50% stiffness reduction in br 1-11 and br 7-17 (dashed lines in Figure 3(a)); 2) DP2B: 25% stiffness reduction in br 1-11 and br 7-17; 3) DP3B: same damage as DP1B, but in addition 25% stiffness reduction in br 19-29 and br 25-35 (dashed lines in Figure 3(b)); 4) DP3Bu: 50% and 25% stiffness reduction in br 1-11 and br 19-29 (dashed lines in Figure 3(c)).

The simulated time-domain data used for structural stiffness identification were generated by a Matlab program for the simulated Phase II Benchmark [36] from broadband ambient-vibration excitations at each floor; simulated noise was added at the measured DOFs with variance equal to 10% of the mean square of the time histories. Two instrumentation scenarios, the full-sensor and partial-sensor scenarios, are considered. For the full-sensor scenario, measurements are available at the center of each side at each floor with the directions parallel to the side in either the $+x$ direction or $+y$ direction. For the partial-sensor scenario, only the measurements at the third floor (mid-height) and the roof are available.

The extraction of the modal properties from the time-domain simulated data was achieved by using the modal identification procedure called MODE-ID [30] and the results were presented in Ching and Beck [16]. For the monitoring stage, a total of ten sets of independent MAP estimates of the experimental modal parameters ($\hat{\boldsymbol{\omega}}_{d,r}^2$ and $\hat{\boldsymbol{\psi}}_{d,r}$, $r = 1, \ldots, 10$) are extracted from ten time segments of equal duration of 20s (10,000 sampling points with sampling frequencies of 500 Hz), and eight modes ($N_m = 8$), four in the strong ($x$) direction and four in the weak ($y$) direction, of the structure are identified for each time segment. For the calibration stage, in order to get more accurate MAP estimates of the stiffness parameters, we change the variable "time_duration" to 2000s in the Matlab program for the simulated Phase II Benchmark and increase the number of time-history segments for tests to one hundred ($N_s = 100$) to identify one hundred sets of modal parameters from the undamaged structure.

To locate the faces sustaining brace damage, the identification model class is defined with the stiffness matrix $\mathbf{K}$ parameterized as

$$\mathbf{K}(\boldsymbol{\theta}) = \mathbf{K}_0 + \sum_s \sum_v \theta_{sv} \overline{\mathbf{K}}_{sv} \qquad (48)$$

where $s = 1, \ldots, 4$ refers to the story number and $v = +x, -x, +y, -y$ indicates the direction of the outward normal of each face at each floor, yielding a stiffness scaling parameter vector $\boldsymbol{\theta}$ with sixteen components, corresponding to sixteen substructures (four faces of four stories). The $\overline{\mathbf{K}}_{sv}$'s are the 'nominal' stiffness matrices to make the nominal value of each stiffness parameter to be 1.0; they are computed based on shear-building assumptions by projecting the full stiffness matrix for the HKUST 120-DOF structural model [36] onto 12 DOFs for a 3-D shear-building model with rigid floors (three DOFs per floor: translations parallel to the x-and y-axes and rotation about the z-axis). $\mathbf{K}_0$ is the nominal stiffness matrix contribution from the columns and beams. The mass matrix is obtained by a similar procedure and is taken as known in the defined model class.

During the calibration stage, we utilize Algorithm 1 based on the 100 independent sets of identified modal parameters from the undamaged structure to find the MAP structural stiffness scaling parameters $\widehat{\boldsymbol{\theta}}_u$ and their corresponding c.o.v (coefficient of variation), which are shown in Tables 1-2 for the two sensor scenarios. The c.o.v of the stiffness scaling parameters is calculated from the ratio of the square root of the diagonal elements of the posterior covariance matrix $\boldsymbol{\Sigma}$ given in (20(b)) to the mean values $\boldsymbol{\mu}$ in (20(a)), and small values of c.o.v demonstrate high confidence (conditional on the modeling) of the identification. It is evident from the tables that the MAP estimates of all stiffness scaling parameters $\theta_j$ are accurate, i.e, errors are smaller than 0.6% and 3.7% for the full sensor and partial sensor scenarios, respectively.

During the monitoring stage, we choose the MAP value $\widehat{\boldsymbol{\theta}}_u$ from the calibration stage as pseudo-data for $\boldsymbol{\theta}$ and perform Algorithm 2 based on ten sets of identified modal parameters as the primary data. The identified stiffness ratios of the MAP stiffness parameters with respect to those for the undamaged case (RB) are tabulated in Tables 1 and 2, for full-sensor and partial-sensor scenarios, respectively, along with the corresponding c.o.v. The actual damaged locations are made bold for comparison. It is found damage patterns are reliably detected in both qualitative and quantitative ways. The identified stiffness ratios for the actual damaged substructures are close to their actual values (0.943 and 0.887) and all components with non-bold font have their stiffness ratio to be exactly equal to one. In addition, the corresponding c.o.v. for those exactly identified non-damaged components is equal to zero, which means these substructures have no stiffness reduction with full confidence compared with that of the calibration stage. This is a benefit of the proposed sparse Bayesian formulation which reduces the uncertainty of the unchanged components. It produces sparse models by learning the hyper-parameter $\boldsymbol{\alpha}$, where $\tilde{\alpha}_{sv} \to 0$ implies that $\boldsymbol{\Sigma}_{(sv)(sv)} \to 0$ and $\tilde{\theta}_{sv} \to (\hat{\theta}_u)_{sv}$.

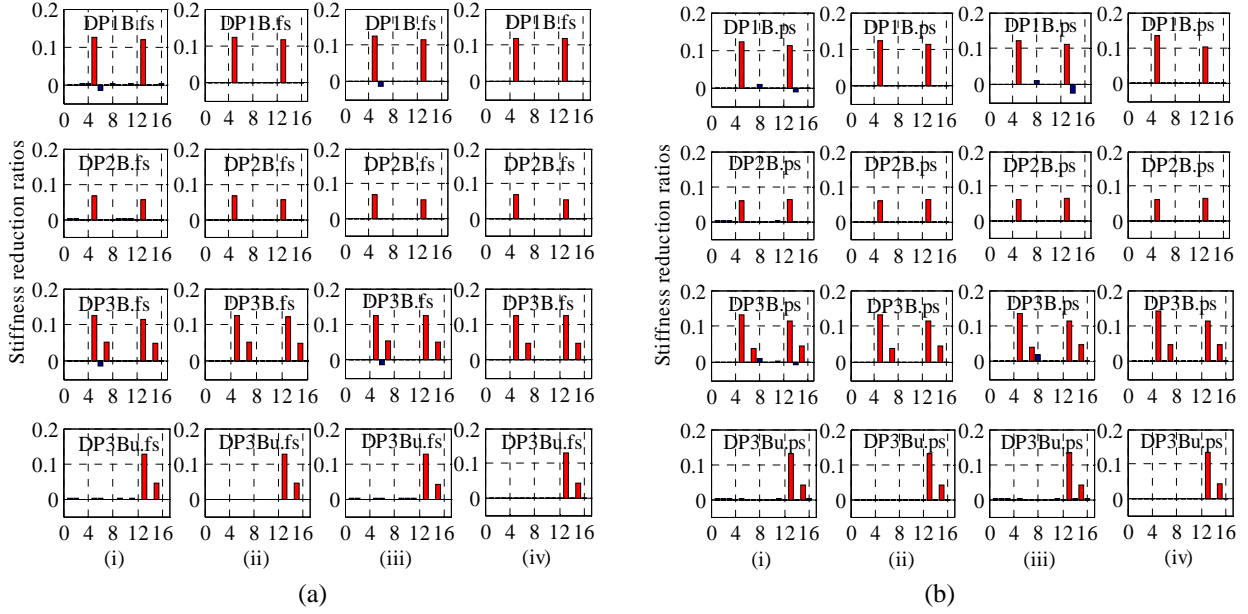

**Fig. 4.** Simulated Phase II benchmark: Comparison of stiffness reduction ratios of two methods for the 16 substructures for the four damage scenarios: (i) The algorithm in [29] with $\lambda = 0$; (ii) The algorithm in [29] with $\lambda$ optimized in each iteration; (iii) Algorithm with no constraints; (iv) Algorithm 2. Two instrumentation scenarios are presented: (a) full-sensor; (b) partial-sensor (red bars indicate actual damage locations).

In order to further portray the identification performance of the proposed algorithms, the identified stiffness reduction ratios of Algorithm 2 are compared with those of our previous Bayesian inversion algorithm in [29] for various damage cases, and the results are presented in Figure 4, where the stiffness reduction ratio is computed as the difference between each MAP value $\left(\hat{\boldsymbol{\theta}}_u\right)_{sv}$ of the stiffness scaling parameters from the calibration stage and the MAP value $\widetilde{\boldsymbol{\theta}}_{sv}$ from the current monitoring stage, normalized by $\left(\hat{\boldsymbol{\theta}}_u\right)_{sv}$. A variant of Algorithm 2 with the stiffness increase constraint removed (Step 6 is omitted) is also studied in order to investigate the effect of this constraint on the identification performance. The results for Algorithm 2 without constraints and Algorithm 2 are presented in Figures 4 (iii) and 4 (iv), respectively, while for the method in [29], the results with $\lambda$ fixed at zero and $\lambda$ optimized are given in Figures 4(i) and 4(ii), respectively. Note that a uniform hyper-prior over $\boldsymbol{\alpha}$ is utilized for Algorithm 2, which is equal to the Laplace prior case with $\lambda = 0$. In this case, all 16 components $\theta_{sv}$ of $\boldsymbol{\theta}$ are updated where $\theta_1,\ldots, \theta_{16}$ correspond to the order listed in the first column of Table 1 and 2. When running Algorithm 2 without constraints, some false damage detections are observed in the partial sensor case although they have only small identified stiffness reductions (Figure 4b(iii)), showing less sparse results. Fortunately, no false detections are observed when we run Algorithm 2, and hence an appropriate sparseness level of stiffness reductions is reliably inferred when imposing the constraint in Step 6 to suppress stiffness increases. For the algorithm in [29], a few false damage alarms are observed in Figure 4b(i) for $\lambda = 0$ but they are effectively suppressed when the Laplace prior with optimal $\lambda$ is used (Figure 4b(ii)); in fact, the latter algorithm has the same perfect damage assessment performance as the proposed Algorithm 2 in the simulated data example but, as shown in the next section, this is no longer true for the real data case.

A more complete picture of substructure damage is given by the probabilities of damage, $P_{sv}^{dam}(f)$ with different severities $f \in [0,1]$, calculated according to Eq. (44). In Figures 5 and 6, the probability of damage for the sixteen $\theta_{uv}'s$ for various brace damage patterns by applying Algorithm 2 are shown, for both full and partial-sensor scenarios, respectively. It is clear in all cases that the substructures with real stiffness losses have damage with a probability of unity. The result for DP1B.fs (Figure 5(a)) match almost perfectly with the real stiffness loss: the means of the damage are estimated as 12.0% for $1, +y$ and $1, -y$ faces, compared to the actual values of 11.3%. For the faces that are undamaged, most of the mean stiffness scaling parameters are close to zero, and the probability is very small when the damage extent exceeds 0.5%, which is within the tolerable uncertainty level for applications. The results obtained for this example are close to those presented in [29] for an earlier version of our theory, although some approximations used there have been removed in the current more rigorous theory.

**TABLE 1:** Simulated Phase II: Identification results from Algorithms 1 and 2 for the full-sensor scenario.

| Para-meter | RB.fs* | | DP1B.fs | | DP2B.fs | | DP3B.fs | | DP3Bu.fs | |
|---|---|---|---|---|---|---|---|---|---|---|
| | MAP value ($\hat{\boldsymbol{\theta}}_u$) | c.o.v. (%) | MAP ratio ($\tilde{\boldsymbol{\theta}}/\hat{\boldsymbol{\theta}}_u$) | c.o.v. (%) | MAP ratio ($\tilde{\boldsymbol{\theta}}/\hat{\boldsymbol{\theta}}_u$) | c.o.v. (%) | MAP ratio ($\tilde{\boldsymbol{\theta}}/\hat{\boldsymbol{\theta}}_u$) | c.o.v. (%) | MAP ratio ($\tilde{\boldsymbol{\theta}}/\hat{\boldsymbol{\theta}}_u$) | c.o.v. (%) |
| $\theta_{1,+x}$ | 1.003 | 0.191 | 1.000 | 0.000 | 1.000 | 0.000 | 1.000 | 0.000 | 1.000 | 0.000 |
| $\theta_{2,+x}$ | 0.999 | 0.098 | 1.000 | 0.000 | 1.000 | 0.000 | 1.000 | 0.000 | 1.000 | 0.000 |
| $\theta_{3,+x}$ | 1.000 | 0.095 | 1.000 | 0.000 | 1.000 | 0.000 | 1.000 | 0.000 | 1.000 | 0.000 |
| $\theta_{4,+x}$ | 1.003 | 0.075 | 1.000 | 0.000 | 1.000 | 0.000 | 1.000 | 0.000 | 1.000 | 0.000 |
| $\theta_{1,+y}$ | 0.999 | 0.240 | **0.881** | **0.196** | **0.933** | **0.189** | **0.875** | **0.201** | 1.000 | 0.000 |
| $\theta_{2,+y}$ | 1.001 | 0.122 | 1.000 | 0.000 | 1.000 | 0.000 | 1.000 | 0.105 | 1.000 | 0.000 |
| $\theta_{3,+y}$ | 1.002 | 0.102 | 1.000 | 0.000 | 1.000 | 0.000 | **0.952** | **0.086** | 1.000 | 0.000 |
| $\theta_{4,+y}$ | 0.998 | 0.099 | 1.000 | 0.000 | 1.000 | 0.000 | 1.000 | 0.000 | 1.000 | 0.000 |
| $\theta_{1,-x}$ | 0.994 | 0.334 | 1.000 | 0.000 | 1.000 | 0.000 | 1.000 | 0.000 | 1.000 | 0.000 |
| $\theta_{2,-x}$ | 0.999 | 0.163 | 1.000 | 0.000 | 1.000 | 0.000 | 1.000 | 0.000 | 1.000 | 0.000 |
| $\theta_{3,-x}$ | 1.002 | 0.169 | 1.000 | 0.000 | 1.000 | 0.000 | 1.000 | 0.000 | 1.000 | 0.000 |
| $\theta_{4,-x}$ | 1.001 | 0.128 | 1.000 | 0.000 | 1.000 | 0.000 | 1.000 | 0.000 | 1.000 | 0.000 |
| $\theta_{1,-y}$ | 1.003 | 0.225 | **0.882** | **0.192** | **0.946** | **0.183** | **0.874** | **0.196** | **0.872** | **0.187** |
| $\theta_{2,-y}$ | 1.005 | 0.118 | 1.000 | 0.000 | 1.000 | 0.000 | 1.000 | 0.000 | 1.000 | 0.000 |
| $\theta_{3,-y}$ | 1.002 | 0.093 | 1.000 | 0.000 | 1.000 | 0.000 | **0.952** | **0.079** | **0.958** | **0.071** |
| $\theta_{4,-y}$ | 0.999 | 0.101 | 1.000 | 0.000 | 1.000 | 0.000 | 1.000 | 0.000 | 1.000 | 0.000 |

*.fs denotes the full-sensor scenario.

**TABLE 2:** Simulated Phase II: Identification results from Algorithms 1 and 2 for the partial-sensor scenario.

| Para-meter | RB.ps* | | DP1B.ps | | DP2B.ps | | DP3B.ps | | DP3Bu.ps | |
|---|---|---|---|---|---|---|---|---|---|---|
| | MAP value ($\hat{\boldsymbol{\theta}}_u$) | c.o.v. (%) | MAP ratio ($\tilde{\boldsymbol{\theta}}/\hat{\boldsymbol{\theta}}_u$) | c.o.v. (%) | MAP ratio ($\tilde{\boldsymbol{\theta}}/\hat{\boldsymbol{\theta}}_u$) | c.o.v. (%) | MAP ratio ($\tilde{\boldsymbol{\theta}}/\hat{\boldsymbol{\theta}}_u$) | c.o.v. (%) | MAP ratio ($\tilde{\boldsymbol{\theta}}/\hat{\boldsymbol{\theta}}_u$) | c.o.v. (%) |
| $\theta_{1,+x}$ | 0.984 | 0.192 | 1.000 | 0.000 | 1.000 | 0.000 | 1.000 | 0.000 | 1.000 | 0.000 |
| $\theta_{2,+x}$ | 1.010 | 0.094 | 1.000 | 0.000 | 1.000 | 0.000 | 1.000 | 0.000 | 1.000 | 0.000 |
| $\theta_{3,+x}$ | 1.026 | 0.094 | 1.000 | 0.000 | 1.000 | 0.000 | 1.000 | 0.000 | 1.000 | 0.000 |
| $\theta_{4,+x}$ | 0.963 | 0.075 | 1.000 | 0.000 | 1.000 | 0.000 | 1.000 | 0.000 | 1.000 | 0.000 |
| $\theta_{1,+y}$ | 0.981 | 0.235 | **0.865** | **0.203** | **0.941** | **0.183** | **0.858** | **0.208** | 1.000 | 0.000 |
| $\theta_{2,+y}$ | 0.992 | 0.122 | 1.000 | 0.000 | 1.000 | 0.000 | 1.000 | 0.000 | 1.000 | 0.000 |
| $\theta_{3,+y}$ | 1.005 | 0.100 | 1.000 | 0.000 | 1.000 | 0.000 | **0.952** | **0.087** | 1.000 | 0.000 |
| $\theta_{4,+y}$ | 1.012 | 0.098 | 1.000 | 0.000 | 1.000 | 0.000 | 1.000 | 0.000 | 1.000 | 0.000 |
| $\theta_{1,-x}$ | 0.997 | 0.325 | 1.000 | 0.000 | 1.000 | 0.000 | 1.000 | 0.000 | 1.000 | 0.000 |
| $\theta_{2,-x}$ | 0.985 | 0.162 | 1.000 | 0.000 | 1.000 | 0.000 | 1.000 | 0.000 | 1.000 | 0.000 |
| $\theta_{3,-x}$ | 0.976 | 0.171 | 1.000 | 0.000 | 1.000 | 0.000 | 1.000 | 0.000 | 1.000 | 0.000 |
| $\theta_{4,-x}$ | 1.021 | 0.125 | 1.000 | 0.000 | 1.000 | 0.000 | 1.000 | 0.000 | 1.000 | 0.000 |
| $\theta_{1,-y}$ | 0.998 | 0.221 | **0.896** | **0.196** | **0.936** | **0.179** | **0.886** | **0.201** | **0.867** | **0.183** |
| $\theta_{2,-y}$ | 0.995 | 0.117 | 1.000 | 0.000 | 1.000 | 0.000 | 1.000 | 0.000 | 1.000 | 0.000 |
| $\theta_{3,-y}$ | 0.988 | 0.093 | 1.000 | 0.000 | 1.000 | 0.000 | **0.954** | **0.081** | **0.957** | **0.069** |
| $\theta_{4,-y}$ | 1.006 | 0.097 | 1.000 | 0.000 | 1.000 | 0.000 | 1.000 | 0.000 | 1.000 | 0.000 |

*.ps denotes the partial-sensor scenario.

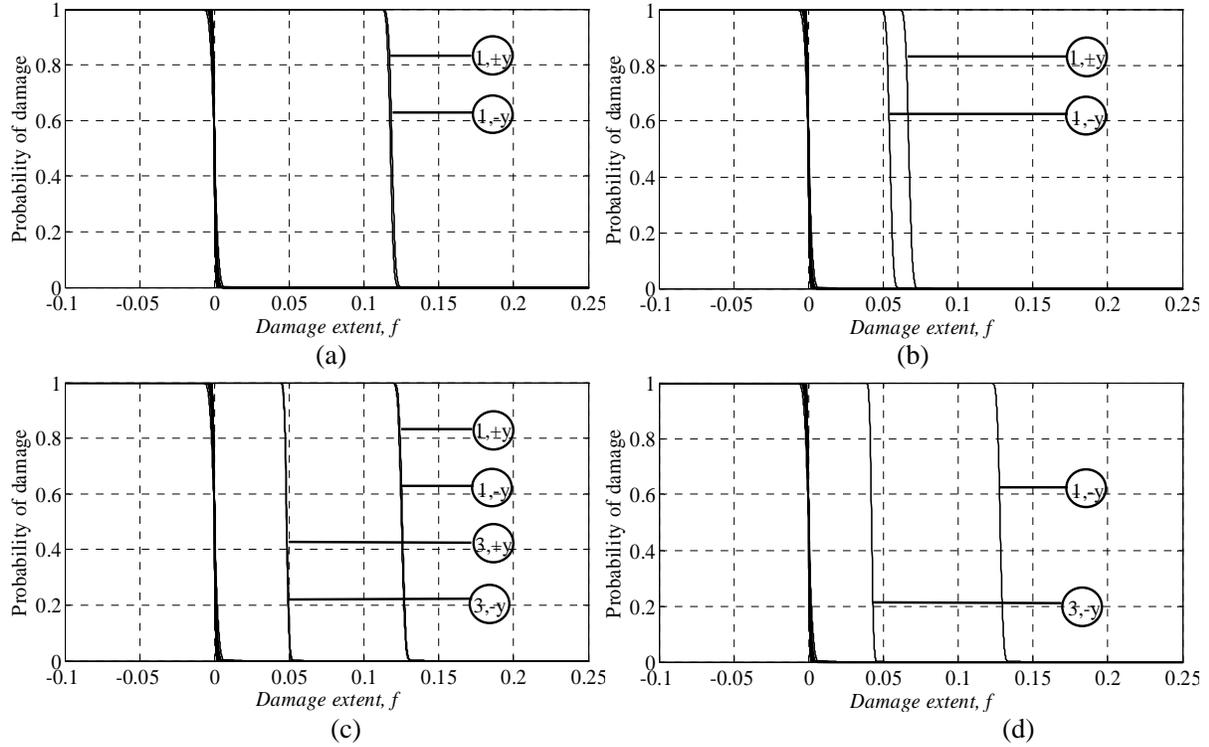

**Fig. 5.** Simulated Phase II benchmark: probability of damage exceeding *f* for the 16 substructures calculated using Algorithm 2 for the full-sensor scenario: (a) DP1B.fs; (b) DP2B.fs; (c) DP3B.fs, and (d) DP3B.fs.

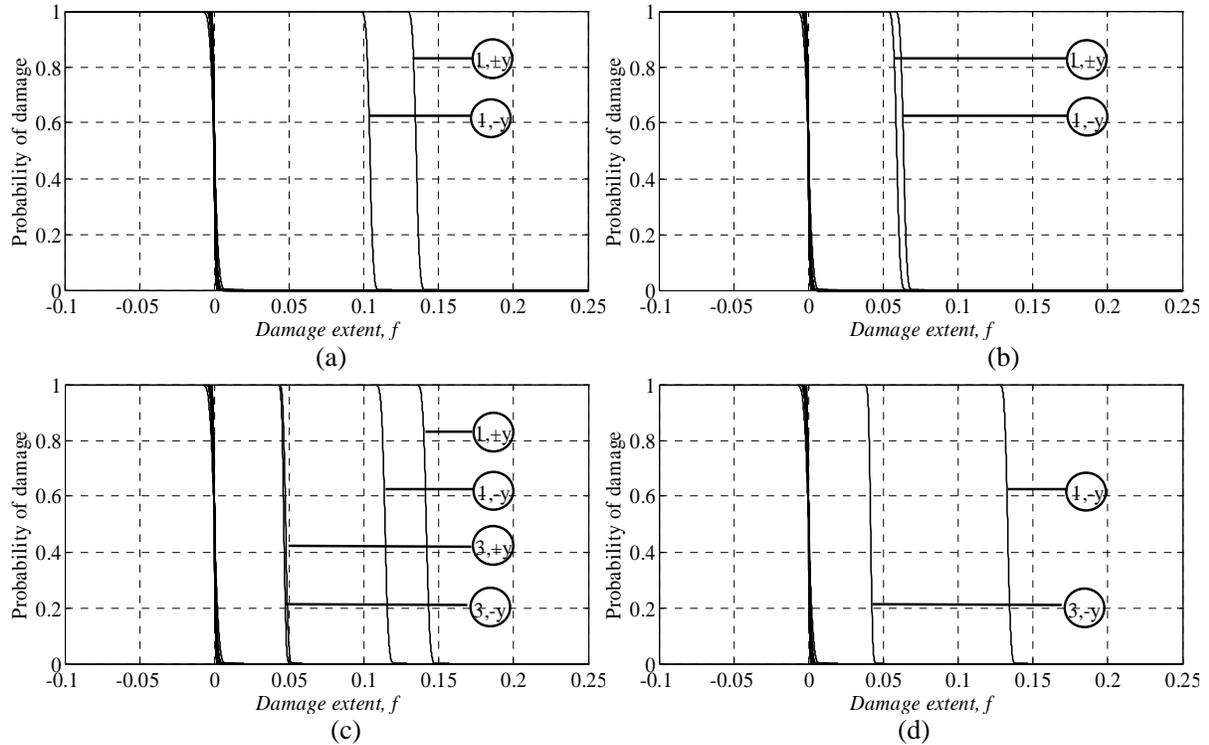

**Fig. 6.** Simulated Phase II benchmark: probability of damage exceeding *f* for the 16 substructures calculated using Algorithm 2 for the partial-sensor scenario: (a) DP1B.ps; (b) DP2B.ps; (c) DP3B.ps, and (d) DP3B.ps.

*5.2 Experimental Phase II Benchmark problem*

In the second example, the proposed methodology is applied to the brace damage cases in the IASC-ASCE experimental Phase II benchmark problem [37]. Six configurations are considered, in which Config 1 is the calibration (undamaged) case and Configs 2-5 are five damage cases: 1) Config 2: removal of all braces (100% stiffness loss) on the –y face; 2) Config 3: removal of the left-hand-side brace in each story on the –y face; 3) Config 4: removal of the left-hand-side braces in the first and fourth stories on the –y face; 4) Config 5: removal of the left hand-side brace in the first story on the –y face; 5) Config 6: removal of two braces in the second story on the $+x$ face. For each configuration, experimental acceleration data measured from the benchmark structure by impact of a sledgehammer are utilized. Sensor data were collected from five Kinemetrics EPI sensors placed near the base and floor centers (Nodes 5, 14, 23, 32, and 41 in Fig. 2) sensing the accelerations in the $+y$ direction, and from ten force–balance accelerometers mounted on the $+y$ and $-y$ faces of the base and all floors (Nodes 2, 8, 11, 17, 20, 26, 29, 35, 38, and 44 in Fig. 2) sensing accelerations in the $+x$ direction.

For each configuration, the time-domain data from each sensor is divided into three segments and three independent sets of experimental modal parameters are extracted by MODE-ID [30]. Five modes ($N_m = 5$), consisting of the first and second translation modes in the $x$ and $y$ directions and the first torsion mode, are identified from each time segment and the results can be found in [16].

A 3D 12-DOF shear building model that was used for the braced cases in simulated Phase II is adapted for inference of the sixteen stiffness scaling parameters (four faces of four stories). The nominal stiffness matrices $\overline{\mathbf{K}}_{uv}$ are identical to those used in the simulated Phase II study. Note that the stiffness losses for a particular face with removal of one and two braces are 22.6% and 45.2%, respectively. This corresponds to a stiffness scaling parameter of 77.4% and 54.9%, respectively, of the undamaged values.

We first run Algorithm 1 to get the MAP estimates of stiffness scaling parameter $\hat{\boldsymbol{\theta}}_u$ in the calibration stage. Notice that the available information of three independent sets of identified modal parameter $\hat{\boldsymbol{\omega}}_u^2$ and $\hat{\boldsymbol{\psi}}_u$ is insufficient to support an accurate inference of the stiffness scaling parameters. We therefore assume that the stiffness scaling parameters of $+x$ and $-x$ faces for all floors, and $+y$ and $-y$ faces for all floors, are identical in order to reduce the number of stiffness model parameters to two. During the monitoring stage, the MAP value $\widehat{\boldsymbol{\theta}}_u$ from the calibration stage is used as pseudo-data for $\boldsymbol{\theta}$ and Algorithm 2 is implemented based on three sets of identified modal parameters $\widehat{\boldsymbol{\omega}}_d^2$ and $\widehat{\boldsymbol{\psi}}_d$; as in Section 5.1, in this case, all 16 components $\theta_j$ of $\boldsymbol{\theta}$ are updated where $\theta_1, \ldots, \theta_{16}$ correspond to the order listed in the first column of Table 3.

Based on the results from Algorithm 2, the stiffness ratios and the associated c.o.v. are tabulated in Table 3. The actual damaged locations have been made bold-faced and it is seen that all brace damage is effectively detected although there is a false detection for $\theta_{16} = \theta_{4,-y}$ of Config. 5 with 3.7% stiffness reduction, which is only 10% of the stiffness reduction value for the actual damaged substructure corresponding to $\theta_{1,-y}$; furthermore, the value for no damage (1.000) is less than one standard deviation (5.2%) from the MAP value of 0.964 for $\theta_{4,-y}$ and so the probability of no damage is quite high. The MAP estimate of stiffness ratios for a particular face with removal of one and two braces are underestimated as approximately 50% and 30%, respectively, compared to the real stiffness ratios of 77.4% and 54.9%, respectively. These results are acceptable since it is much more important to make accurate localizations of significant damage than to estimate the damage extent accurately in real applications. For posterior uncertainty quantification, the c.o.v. is generally much higher compared with those values in Tables 1-2 for the simulated benchmark due to larger modeling errors in this real case, especially for those components corresponding to real damage locations.

In Figure 7, the proposed Algorithm 2 without constraints (Figure 7(c)) and Algorithm 2 (Figure 7(d)) are compared with the method in [29] by presenting the stiffness reduction ratios for various damage cases, where hyper-parameter $\lambda = 0$ (Figure 7(a)) and $\lambda$ optimized for the assigned Laplace prior (Figure 7(b)) are considered for the method in [29]. It is interesting to see that totally over-sparse stiffness reduction models with no nonzero components are produced when $\lambda$ is optimized, which is consistent with the discussion in Remark 4.4. In comparison with the method in [29] with $\lambda = 0$, Algorithm 2 with no constraints has higher sparseness (Figure 7 (c)). Nevertheless, false detections are still observed in $\theta_{12} = \theta_{4,-x}$ for Config 4 and $\theta_{16} = \theta_{4,-y}$ for Config 7 with significant identified stiffness reduction for Algorithm 2 without constraints. In addition, some undamaged substructures show a significant increase in stiffness that is unrealistic. Fortunately, incorporating the additional constraint to suppress stiffness increases results in Algorithm 2 localizing all stiffness reductions exactly (Figure 7 (d)).

**TABLE 3:** Stiffness ratios obtained from Algorithms 1 and 2 for the braced-damage hammer-impact cases in experimental Phase II.

| Parameter | Config1 MAP value ($\widehat{\boldsymbol{\theta}}_u/\widehat{\boldsymbol{\theta}}_u$) | c.o.v. (%) | Config2 MAP ratio ($\widetilde{\boldsymbol{\theta}}/\widehat{\boldsymbol{\theta}}_u$) | c.o.v. (%) | Config3 MAP ratio ($\widetilde{\boldsymbol{\theta}}/\widehat{\boldsymbol{\theta}}_u$) | c.o.v. (%) | Config4 MAP ratio ($\widetilde{\boldsymbol{\theta}}/\widehat{\boldsymbol{\theta}}_u$) | c.o.v. (%) | Config5 MAP ratio ($\widetilde{\boldsymbol{\theta}}/\widehat{\boldsymbol{\theta}}_u$) | c.o.v. (%) | Config6 MAP ratio ($\widetilde{\boldsymbol{\theta}}/\widehat{\boldsymbol{\theta}}_u$) | c.o.v. (%) |
|---|---|---|---|---|---|---|---|---|---|---|---|---|
| $\theta_{1,+x}$ | 1.000 | 4.509 | 1.000 | 0.000 | 1.000 | 0.000 | 1.000 | 0.000 | 1.000 | 0.000 | 1.000 | 0.000 |
| $\theta_{2,+x}$ | 1.000 | 9.658 | 1.000 | 0.000 | 1.000 | 0.000 | 1.000 | 0.000 | 1.000 | 0.000 | **0.835** | **15.719** |
| $\theta_{3,+x}$ | 1.000 | 4.509 | 1.000 | 0.000 | 1.000 | 0.000 | 1.000 | 0.000 | 1.000 | 0.000 | 1.000 | 0.000 |
| $\theta_{4,+x}$ | 1.000 | 9.658 | 1.000 | 0.000 | 1.000 | 0.000 | 1.000 | 0.000 | 1.000 | 0.000 | 1.000 | 0.000 |
| $\theta_{1,+y}$ | 1.000 | 4.509 | 1.000 | 0.000 | 1.000 | 1.593 | 1.000 | 0.000 | 1.000 | 0.000 | 0.994 | 3.458 |
| $\theta_{2,+y}$ | 1.000 | 9.658 | 1.000 | 0.000 | 1.000 | 0.000 | 1.000 | 1.981 | 0.999 | 1.345 | 1.000 | 0.000 |
| $\theta_{3,+y}$ | 1.000 | 4.509 | 1.000 | 0.000 | 1.000 | 0.000 | 1.000 | 0.000 | 1.000 | 0.000 | 1.000 | 0.000 |
| $\theta_{4,+y}$ | 1.000 | 9.658 | 1.000 | 0.000 | 1.000 | 0.000 | 1.000 | 0.000 | 1.000 | 0.000 | 1.000 | 0.000 |
| $\theta_{1,-x}$ | 1.000 | 4.509 | 1.000 | 0.000 | 1.000 | 0.000 | 1.000 | 0.000 | 1.000 | 0.000 | 1.000 | 0.000 |
| $\theta_{2,-x}$ | 1.000 | 9.658 | 1.000 | 0.000 | 1.000 | 0.000 | 1.000 | 0.000 | 1.000 | 0.000 | 1.000 | 0.000 |
| $\theta_{3,-x}$ | 1.000 | 4.509 | 1.000 | 0.000 | 1.000 | 0.000 | 1.000 | 0.000 | 1.000 | 0.000 | 1.000 | 0.000 |
| $\theta_{4,-x}$ | 1.000 | 9.658 | 1.000 | 0.000 | 1.000 | 0.000 | 1.000 | 0.000 | 1.000 | 0.000 | 1.000 | 0.000 |
| $\theta_{1,-y}$ | 1.000 | 4.509 | **0.302** | **36.148** | **0.443** | **28.930** | **0.767** | **22.028** | **0.677** | **13.541** | 1.000 | 0.000 |
| $\theta_{2,-y}$ | 1.000 | 9.658 | **0.370** | **58.243** | **0.646** | **34.911** | 1.000 | 0.000 | 1.000 | 0.000 | 1.000 | 0.000 |
| $\theta_{3,-y}$ | 1.000 | 4.509 | **0.268** | **32.306** | **0.386** | **25.493** | 1.000 | 1.359 | 1.000 | 0.000 | 1.000 | 0.000 |
| $\theta_{4,-y}$ | 1.000 | 9.658 | **0.160** | **45.287** | **0.293** | **33.986** | **0.608** | **22.014** | 0.964 | 5.245 | 1.000 | 0.000 |

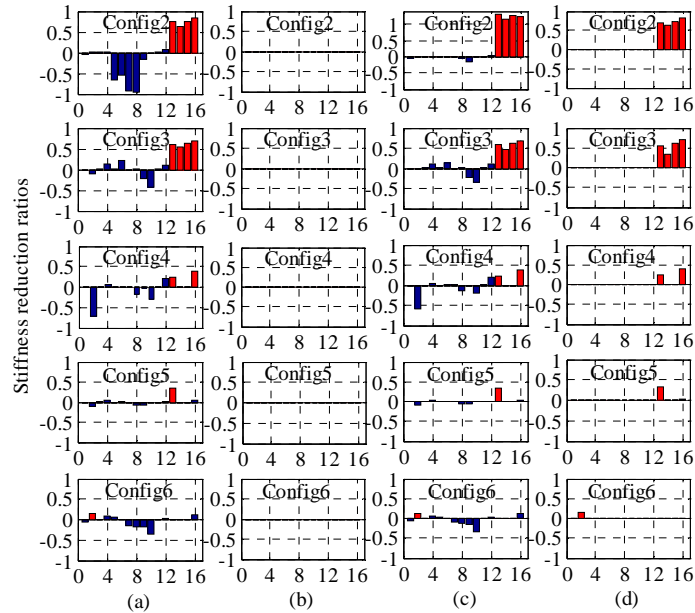

**Fig. 7.** Experimental Phase II benchmark: comparison of stiffness reduction ratios from two Bayesian inversion methods for the 16 substructures for five brace-damage scenarios: (a) The algorithm in [29] with $\lambda = 0$; (b) The algorithm in [29] with $\lambda$ optimized in each iteration; (c) Algorithm 2 with no constraints; (d) Algorithm 2 (Note that a negative value corresponds to a stiffness gain).

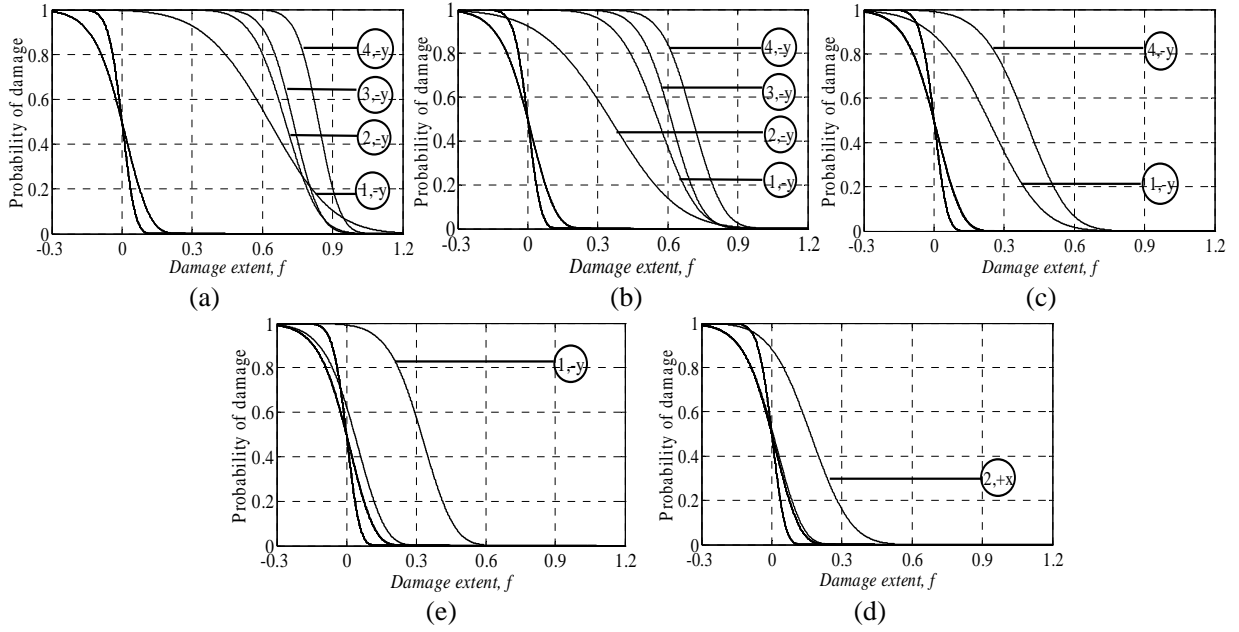

**Fig. 8.** Probability of damage exceeding *f* for the 16 substructures using Algorithm 2, for the brace-damage cases in experimental Phase II benchmark: (a) Config 2; (b) Config 3; (c) Config 4; (d) Config 5 and (e) Config 6.

The damage probability curves for the sixteen stiffness scaling parameters **θ** are plotted in Fig. 8 for Configs. 2–6. All the actual damaged substructures are clearly shown to have a large damage probability when their stiffness scaling parameter has been reduced by the actual fraction. For the undamaged substructures, the posterior medians of the estimated damage corresponding to a damage probability of 0.5 are around zero, although in some cases small probabilities of about 0.2 are estimated for damage fractions exceeding 10%, which is due to the large posterior uncertainty for the stiffness scaling parameters in the calibration and monitoring stages (Table 3).

## 6. Concluding remarks

In the paper, we examine the problem of inferring damage-induced stiffness reductions in a structure from SHM data by taking a hierarchical Bayesian model perspective, which leads to a new sparse Bayesian learning algorithm for probabilistic damage detection with noisy incomplete modal data. This framework has the following five attractive features. First, for each chosen substructure, not only the most probable estimate of the spatially-sparse stiffness changes based on the identified modal parameters is inferred but also the associated posterior uncertainties of the stiffness parameters are quantified, including the probability of substructure damage of various amounts. Second, in addition to the prior information that structural damage is spatially sparse, we incorporate the constraint that the stiffness parameters cannot have values in a damaged state that are greater those of the calibration state, which reduces the occurrence of incorrect identifications of spatially-sparse stiffness changes. Third, instead of solving the nonlinear eigenvalue problem of a structural model, the eigenvalue equations are used as a soft constraint via a prior distribution, which avoids a substantial computational cost. Fourth, all uncertain hyper-parameters in the proposed hierarchical Bayesian model are estimated solely from the identified modal data, giving an algorithm for which no user-intervention is needed to set algorithm parameters, unlike most other methods. Fifth, no explicit matching of model and experimental modes is needed, thereby avoiding a practical difficulty in model updating using identified modal parameters.

The superior effectiveness and robustness of the proposed method compared with two previously-published Bayesian methods [18, 29] was successfully demonstrated through the analysis of the IASC–ASCE Phase II simulated and experimental benchmark problems for identification of brace damage of the benchmark structure. The results show that our damage assessment method is very reliable, despite significant modelling error in the structural model in the real experiments. In all test cases, there were essentially no false or missed alerts, that is, false positives and negatives.

For future studies, it would be useful to explore the uncertainties in both the identified system modal parameters and the stiffness parameters from the calibration stage, since these are effectively ignored when only their MAP values are utilized.

**Acknowledgements**

This work was supported by the NSF (Award No. EAR-0941374) and the NSFC (Grant No. 51308161).

**Nomenclature**

$N_m$ = Number of extracted modes in the modal identification ($i = 1, \dots, N_m$)

$N_o$ = Number of measured degrees of freedom

$N_d$ = Number of degrees of freedom of the identification model ($k = 1, \dots, N_d$)

$N_s$ = Number of time segments of measured modal data ($r = 1, \dots, N_s$)

$N_\theta$ = Number of substructures considered ($j = 1, \ldots, N_\theta$)

$\mathbf{M}, \mathbf{K}$ = Mass and stiffness matrices

$\boldsymbol{\theta}$ = Structural stiffness scaling parameters

$\boldsymbol{\phi}_r, \omega_r^2$ = System mode shape and system natural frequencies of the $rth$ mode

$\beta$ = Equation error precision parameter

$\hat{\omega}_{r,i}^2, \hat{\boldsymbol{\psi}}_{r,i}$ = MAP estimates of modal frequency and mode shape of $i^{th}$ mode from the $r^{th}$ data segment from modal identification

$\hat{\boldsymbol{\theta}}_u$ = MAP estimate of $\boldsymbol{\theta}$ determined from the calibration test data.

$\boldsymbol{\Gamma}$ = Matrix that picks the measured degrees of freedom from the system mode shape

$\mathbf{T}$ = Transformation matrix between the vector of $q$ sets of identified natural frequencies $\hat{\boldsymbol{\omega}}^2$ and the system natural frequencies $\boldsymbol{\omega}^2$

$\alpha$ = Variance parameter for the likelihood function of structural stiffness scaling parameters $\boldsymbol{\theta}$

$\eta, \boldsymbol{\rho}$ = Measurement-error precision parameters for mode shapes and natural frequencies

$\nu, \boldsymbol{\tau}, \kappa$ = Rate parameters controlling the exponential prior distribution of $\eta, \boldsymbol{\rho}, b_0$.

$\boldsymbol{\delta}$ = Set of all hyper-parameters $\boldsymbol{\omega}^2, \boldsymbol{\rho}, \boldsymbol{\tau}, \boldsymbol{\phi}, \eta, \nu, \boldsymbol{\zeta}, \beta, a_0, b_0, \kappa$

**Appendix: MAP estimation of $\boldsymbol{\delta}$**

Iterative equations for calculating the MAP estimate of $\boldsymbol{\delta} = [(\boldsymbol{\omega}^2)^T, \boldsymbol{\rho}^T, \boldsymbol{\tau}^T, \boldsymbol{\phi}^T, \eta, \nu, \boldsymbol{\zeta}^T, \beta, a_0, b_0, \kappa]^T$ are obtained by differentiation of the logarithm of $J_1(\boldsymbol{\delta})$ defined in (25) with respect to each part of $\boldsymbol{\delta}$ and then setting these derivatives to zero. Ignoring terms in the logarithm that are independent of $\boldsymbol{\delta}$, the objective function is defined as:

$$J(\boldsymbol{\delta}) = -\frac{1}{2}(\hat{\boldsymbol{\psi}} - \boldsymbol{\Gamma}\boldsymbol{\phi})^T \mathbf{C}^{-1}(\hat{\boldsymbol{\psi}} - \boldsymbol{\Gamma}\boldsymbol{\phi}) - \frac{1}{2}(\hat{\boldsymbol{\omega}}^2 - \mathbf{L}\boldsymbol{\omega}^2)^T \mathbf{E}^{-1}(\hat{\boldsymbol{\omega}}^2 - \mathbf{L}\boldsymbol{\omega}^2) + \frac{1}{2}(N_d N_m - N_\theta)\log\beta - \frac{1}{2}\log|\mathbf{H}^T\mathbf{H}|$$

$$-\frac{\beta}{2}(\mathbf{b}^T\mathbf{b} - \mathbf{b}^T\mathbf{H}(\mathbf{H}^T\mathbf{H})^{-1}\mathbf{H}^T\mathbf{b}) + \frac{1}{2}N_o N_s N_m \log\eta + \frac{1}{2}N_s \sum_{i=1}^{N_m}(\log\rho_i) - \frac{1}{2}\log|\mathbf{A} + \beta^{-1}(\mathbf{H}^T\mathbf{H})^{-1}|$$

$$-\frac{1}{2}\left(\hat{\boldsymbol{\theta}}_u - (\mathbf{H}^T\mathbf{H})^{-1}\mathbf{H}^T\mathbf{b}\right)^T (\mathbf{A} + \beta^{-1}(\mathbf{H}^T\mathbf{H})^{-1})^{-1}\left(\hat{\boldsymbol{\theta}}_u - (\mathbf{H}^T\mathbf{H})^{-1}\mathbf{H}^T\mathbf{b}\right) + \log\nu - \nu\eta + \sum_{i=1}^{N_m}[\log\tau_i - \tau_i\rho_i]$$

$$+(a_0 - 1)\log\beta - b_0^{-1}\beta - a_0 \log b_0 - \log\Gamma(a_0) + \log\kappa - \kappa b_0 \tag{49}$$

Using the determinant identity, the term $\log|\mathbf{A} + (\beta\mathbf{H}^T\mathbf{H})^{-1}|$ is expressed as:

$$\log|\mathbf{A} + (\beta\mathbf{H}^T\mathbf{H})^{-1}| = \log(|(\beta\mathbf{H}^T\mathbf{H})^{-1}||\beta\mathbf{A}\mathbf{H}^T\mathbf{H} + \mathbf{I}_n|)$$

$$= -N_\theta \log\beta - \log|\mathbf{H}^T\mathbf{H}| + \log|\mathbf{A}| + \log|\mathbf{\Sigma}^{-1}| \tag{50}$$

where $\mathbf{\Sigma} = (\beta\mathbf{H}^T\mathbf{H} + \mathbf{A}^{-1})^{-1}$ is the covariance matrix of $\boldsymbol{\theta}$ defined in (20(b)) and it is readily evaluated for arbitrary values of parameters $\boldsymbol{\phi}$, $\boldsymbol{\alpha}$ and $\beta$.

For the term involving $(\mathbf{A} + (\beta\mathbf{H}^T\mathbf{H})^{-1})^{-1}$ in (49), we first use the Woodbury matrix identity on the inverse:

$$(\mathbf{A} + (\beta\mathbf{H}^T\mathbf{H})^{-1})^{-1} = \beta\mathbf{H}^T\mathbf{H} - \beta\mathbf{H}^T\mathbf{H}(\mathbf{A}^{-1} + \beta\mathbf{H}^T\mathbf{H})^{-1}\beta\mathbf{H}^T\mathbf{H} = \beta\mathbf{H}^T\mathbf{H} - \beta^2\mathbf{H}^T\mathbf{H}\mathbf{\Sigma}\mathbf{H}^T\mathbf{H} \tag{51}$$

and then using $\mathbf{\Sigma}^{-1} = \mathbf{A}^{-1} + \beta\mathbf{H}^T\mathbf{H}$ and $\boldsymbol{\mu} = \mathbf{\Sigma}(\beta\mathbf{H}^T\mathbf{b} + \mathbf{A}^{-1}\hat{\boldsymbol{\theta}}_u)$, we can simplify the term:

$$(\hat{\boldsymbol{\theta}}_u - (\mathbf{H}^T\mathbf{H})^{-1}\mathbf{H}^T\mathbf{b})^T(\mathbf{A} + (\beta\mathbf{H}^T\mathbf{H})^{-1})^{-1}(\hat{\boldsymbol{\theta}}_u - (\mathbf{H}^T\mathbf{H})^{-1}\mathbf{H}^T\mathbf{b})$$

$$= (\hat{\boldsymbol{\theta}}_u - (\mathbf{H}^T\mathbf{H})^{-1}\mathbf{H}^T\mathbf{b})^T(\beta\mathbf{H}^T\mathbf{H} - \beta^2\mathbf{H}^T\mathbf{H}\mathbf{\Sigma}\mathbf{H}^T\mathbf{H})(\hat{\boldsymbol{\theta}}_u - (\mathbf{H}^T\mathbf{H})^{-1}\mathbf{H}^T\mathbf{b})$$

$$= \beta(\hat{\boldsymbol{\theta}}_u - (\mathbf{H}^T\mathbf{H})^{-1}\mathbf{H}^T\mathbf{b})^T\mathbf{H}^T\mathbf{H}(\hat{\boldsymbol{\theta}}_u - (\mathbf{H}^T\mathbf{H})^{-1}\mathbf{H}^T\mathbf{b})$$

$$-\beta^2(\hat{\boldsymbol{\theta}}_u - (\mathbf{H}^T\mathbf{H})^{-1}\mathbf{H}^T\mathbf{b})^T\mathbf{H}^T\mathbf{H}\mathbf{\Sigma}\mathbf{H}^T(\mathbf{H}\hat{\boldsymbol{\theta}}_u - \mathbf{H}(\mathbf{H}^T\mathbf{H})^{-1}\mathbf{H}^T\mathbf{b})$$

$$= \beta(\hat{\boldsymbol{\theta}}_u - (\mathbf{H}^T\mathbf{H})^{-1}\mathbf{H}^T\mathbf{b})^T\mathbf{H}^T\mathbf{H}\left(\hat{\boldsymbol{\theta}}_u - (\mathbf{H}^T\mathbf{H})^{-1}\mathbf{H}^T\mathbf{b} - \beta\mathbf{\Sigma}(\mathbf{H}^T\mathbf{H}\hat{\boldsymbol{\theta}}_u - \mathbf{H}^T\mathbf{b})\right)$$

$$= \beta(\hat{\boldsymbol{\theta}}_u - (\mathbf{H}^T\mathbf{H})^{-1}\mathbf{H}^T\mathbf{b})^T\mathbf{H}^T\mathbf{H}\left(\hat{\boldsymbol{\theta}}_u - (\mathbf{H}^T\mathbf{H})^{-1}\mathbf{H}^T\mathbf{b} - \mathbf{\Sigma}(\beta\mathbf{H}^T\mathbf{H}\hat{\boldsymbol{\theta}}_u + \mathbf{A}^{-1}\hat{\boldsymbol{\theta}}_u - \mathbf{\Sigma}^{-1}\boldsymbol{\mu})\right)$$

$$= \beta(\hat{\boldsymbol{\theta}}_u - (\mathbf{H}^T\mathbf{H})^{-1}\mathbf{H}^T\mathbf{b})^T\mathbf{H}^T\mathbf{H}\left(\hat{\boldsymbol{\theta}}_u - (\mathbf{H}^T\mathbf{H})^{-1}\mathbf{H}^T\mathbf{b} - (\hat{\boldsymbol{\theta}}_u - \boldsymbol{\mu})\right)$$

$$= \beta\left(\hat{\boldsymbol{\theta}}_u - (\mathbf{H}^T\mathbf{H})^{-1}\mathbf{H}^T\mathbf{b} - (\hat{\boldsymbol{\theta}}_u - \boldsymbol{\mu})\right)^T\mathbf{H}^T\mathbf{H}\left(\hat{\boldsymbol{\theta}}_u - (\mathbf{H}^T\mathbf{H})^{-1}\mathbf{H}^T\mathbf{b} - (\hat{\boldsymbol{\theta}}_u - \boldsymbol{\mu})\right)$$

$$+\beta(\mathbf{H}^T\mathbf{H}\hat{\boldsymbol{\theta}}_u - \mathbf{H}^T\mathbf{b})^T(\hat{\boldsymbol{\theta}}_u - \boldsymbol{\mu}) - \beta(\hat{\boldsymbol{\theta}}_u - \boldsymbol{\mu})^T\mathbf{H}^T\mathbf{H}(\hat{\boldsymbol{\theta}}_u - \boldsymbol{\mu})$$

$$= \beta\left(\hat{\boldsymbol{\theta}}_u - (\mathbf{H}^T\mathbf{H})^{-1}\mathbf{H}^T\mathbf{b} - (\hat{\boldsymbol{\theta}}_u - \boldsymbol{\mu})\right)^T\mathbf{H}^T\mathbf{H}\left(\hat{\boldsymbol{\theta}}_u - (\mathbf{H}^T\mathbf{H})^{-1}\mathbf{H}^T\mathbf{b} - (\hat{\boldsymbol{\theta}}_u - \boldsymbol{\mu})\right)$$

$$+(\beta\mathbf{H}^T\mathbf{H}\hat{\boldsymbol{\theta}}_u + \mathbf{A}^{-1}\hat{\boldsymbol{\theta}}_u - \mathbf{\Sigma}^{-1}\boldsymbol{\mu})^T(\hat{\boldsymbol{\theta}}_u - \boldsymbol{\mu}) - \beta(\hat{\boldsymbol{\theta}}_u - \boldsymbol{\mu})^T\mathbf{H}^T\mathbf{H}(\hat{\boldsymbol{\theta}}_u - \boldsymbol{\mu})$$

$$= \beta\left(\hat{\boldsymbol{\theta}}_u - (\mathbf{H}^T\mathbf{H})^{-1}\mathbf{H}^T\mathbf{b} - (\hat{\boldsymbol{\theta}}_u - \boldsymbol{\mu})\right)^T\mathbf{H}^T\mathbf{H}\left(\hat{\boldsymbol{\theta}}_u - (\mathbf{H}^T\mathbf{H})^{-1}\mathbf{H}^T\mathbf{b} - (\hat{\boldsymbol{\theta}}_u - \boldsymbol{\mu})\right)$$

$$+(\hat{\boldsymbol{\theta}}_u - \boldsymbol{\mu})^T\mathbf{\Sigma}^{-1}(\hat{\boldsymbol{\theta}}_u - \boldsymbol{\mu}) - \beta(\hat{\boldsymbol{\theta}}_u - \boldsymbol{\mu})^T\mathbf{H}^T\mathbf{H}(\hat{\boldsymbol{\theta}}_u - \boldsymbol{\mu})$$

$$= \beta(\boldsymbol{\mu} - (\mathbf{H}^T\mathbf{H})^{-1}\mathbf{H}^T\mathbf{b})^T\mathbf{H}^T\mathbf{H}(\boldsymbol{\mu} - (\mathbf{H}^T\mathbf{H})^{-1}\mathbf{H}^T\mathbf{b}) + (\hat{\boldsymbol{\theta}}_u - \boldsymbol{\mu})^T\mathbf{A}^{-1}(\hat{\boldsymbol{\theta}}_u - \boldsymbol{\mu}) \tag{52}$$

Now consider the sum of two of the terms in (49):

$$\beta(\mathbf{b}^T\mathbf{b} - \mathbf{b}^T\mathbf{H}(\mathbf{H}^T\mathbf{H})^{-1}\mathbf{H}^T\mathbf{b}) + (\hat{\boldsymbol{\theta}}_u - (\mathbf{H}^T\mathbf{H})^{-1}\mathbf{H}^T\mathbf{b})^T(\mathbf{A} + \beta^{-1}(\mathbf{H}^T\mathbf{H})^{-1})^{-1}(\hat{\boldsymbol{\theta}}_u - (\mathbf{H}^T\mathbf{H})^{-1}\mathbf{H}^T\mathbf{b})$$

$$= \beta(\mathbf{b}^T\mathbf{b} - \mathbf{b}^T\mathbf{H}(\mathbf{H}^T\mathbf{H})^{-1}\mathbf{H}^T\mathbf{b}) + \beta(\boldsymbol{\mu} - (\mathbf{H}^T\mathbf{H})^{-1}\mathbf{H}^T\mathbf{b})^T\mathbf{H}^T\mathbf{H}(\boldsymbol{\mu} - (\mathbf{H}^T\mathbf{H})^{-1}\mathbf{H}^T\mathbf{b}) + (\hat{\boldsymbol{\theta}}_u - \boldsymbol{\mu})^T\mathbf{A}^{-1}(\hat{\boldsymbol{\theta}}_u - \boldsymbol{\mu})$$

$$= \beta\|\mathbf{H}\boldsymbol{\mu} - \mathbf{b}\|^2 + (\hat{\boldsymbol{\theta}}_u - \boldsymbol{\mu})^T\mathbf{A}^{-1}(\hat{\boldsymbol{\theta}}_u - \boldsymbol{\mu}) \tag{53}$$

Using (50) and (53), we can simplify $J(\boldsymbol{\delta})$ in (49) to:

$$J(\boldsymbol{\delta}) = -\frac{1}{2}(\hat{\boldsymbol{\psi}} - \boldsymbol{\Gamma}\boldsymbol{\phi})^T\mathbf{C}^{-1}(\hat{\boldsymbol{\psi}} - \boldsymbol{\Gamma}\boldsymbol{\phi}) - \frac{1}{2}(\hat{\boldsymbol{\omega}}^2 - \mathbf{L}\boldsymbol{\omega}^2)^T\mathbf{E}^{-1}(\hat{\boldsymbol{\omega}}^2 - \mathbf{L}\boldsymbol{\omega}^2) - \frac{1}{2}(\hat{\boldsymbol{\theta}}_u - \boldsymbol{\mu})^T\mathbf{A}^{-1}(\hat{\boldsymbol{\theta}}_u - \boldsymbol{\mu})$$

$$-\frac{1}{2}\log|\boldsymbol{\Sigma}^{-1}| - \frac{1}{2}\log|\mathbf{A}| - \frac{\beta}{2}\|\mathbf{H}\boldsymbol{\mu} - \mathbf{b}\|^2 + \frac{1}{2}(N_d N_m + 2a_0 - 2)\log\beta + \frac{1}{2}N_o N_s N_m \log\eta$$

$$+\frac{1}{2}N_s \sum_{i=1}^{N_m}(\log\rho_i) + \log\nu - \nu\eta + \sum_{i=1}^{N_m}[\log\tau_i - \tau_i\rho_i] + \sum_{i=1}^{N_\theta}[\log(\lambda_j\zeta_j) - (\lambda_j\alpha_j + \zeta_j\lambda_j)]$$

$$+(a_0 - 1)\log\beta - b_0^{-1}\beta - a_0 \log b_0 - \log\Gamma(a_0) + \log\kappa - \kappa b_0 \tag{54}$$

The derivatives of $\log|\boldsymbol{\Sigma}^{-1}|$ with respect to $\boldsymbol{\phi}$, $\boldsymbol{\alpha}$ and $\beta$ are now derived. To differentiate $\log|\boldsymbol{\Sigma}^{-1}|$ with respect to $\boldsymbol{\phi}$, we first rewrite $\mathbf{H}$ as the following form:

$$\mathbf{H} = \begin{bmatrix} \mathbf{K}_1\boldsymbol{\phi}_1 & \cdots & \mathbf{K}_{N_\theta}\boldsymbol{\phi}_1 \\ \vdots & \ddots & \vdots \\ \mathbf{K}_1\boldsymbol{\phi}_{N_m} & \cdots & \mathbf{K}_{N_\theta}\boldsymbol{\phi}_{N_m} \end{bmatrix}_{N_m N_d \times N_\theta} = \sum_{q=1}^{N_m N_d} \phi_q \boldsymbol{\Pi}_q \tag{55}$$

where $\phi_q$ are the $N_m N_d$ components of $\boldsymbol{\phi} = [\boldsymbol{\phi}_1^T, \ldots, \boldsymbol{\phi}_{N_m}^T]^T \in \mathbb{R}^{N_m N_d \times 1}$ and

$$\boldsymbol{\Pi}_q = \frac{\partial \mathbf{H}}{\partial \phi_q}, q = 1, \ldots, N_m N_d \tag{56}$$

Then:

$$\frac{\partial(\mathbf{H}^T\mathbf{H})}{\partial \phi_q} = \boldsymbol{\Pi}_q^T \mathbf{H} + \mathbf{H}^T \boldsymbol{\Pi}_q = \boldsymbol{\Pi}_q^T \cdot \sum_{s=1}^{N_m N_d} \phi_s \boldsymbol{\Pi}_s + \sum_{s=1}^{N_m N_d} \phi_s \boldsymbol{\Pi}_s^T \cdot \boldsymbol{\Pi}_q$$

$$= 2\phi_q \boldsymbol{\Pi}_q^T \boldsymbol{\Pi}_q + \boldsymbol{\Pi}_q^T \cdot \left(\sum_{s \neq q}^{N_m N_d} \phi_s \boldsymbol{\Pi}_s\right) + \left(\sum_{s \neq q}^{N_m N_d} \phi_s \boldsymbol{\Pi}_s^T\right) \cdot \boldsymbol{\Pi}_q = 2\phi_q \mathbf{T}_q + \mathbf{U}_q + \mathbf{U}_q^T \tag{57}$$

where $\mathbf{T}_q = \boldsymbol{\Pi}_q^T \boldsymbol{\Pi}_q$ and $\mathbf{U}_q = \boldsymbol{\Pi}_q^T \cdot \left(\sum_{s \neq q}^{N_m N_d} \phi_s \boldsymbol{\Pi}_s\right)$ as defined in (30).

Therefore, using the Jacobi's determinant derivative formula:

$$\frac{\partial(\log|\boldsymbol{\Sigma}^{-1}|)}{\partial \phi_q} = \beta \operatorname{tr}\left(\boldsymbol{\Sigma} \cdot \frac{\partial(\mathbf{H}^T\mathbf{H})}{\partial \phi_q}\right) = 2\beta\phi_q \operatorname{tr}(\boldsymbol{\Sigma}\mathbf{T}_q) + \beta \operatorname{tr}(\boldsymbol{\Sigma}\mathbf{U}_q) + \beta \operatorname{tr}(\boldsymbol{\Sigma}\mathbf{U}_q^T)$$

$$= 2\beta\phi_q \operatorname{tr}(\boldsymbol{\Sigma}\mathbf{T}_q) + \beta \operatorname{tr}(\boldsymbol{\Sigma}\mathbf{U}_q) + \beta \operatorname{tr}(\mathbf{U}_q\boldsymbol{\Sigma})$$

$$= 2\beta\phi_q \operatorname{tr}(\boldsymbol{\Sigma}\mathbf{T}_q) + 2\beta \operatorname{tr}(\boldsymbol{\Sigma}\mathbf{U}_q) \tag{58}$$

Consequently, the derivative of $\log|\mathbf{\Sigma}^{-1}|$ with respect to vector $\boldsymbol{\phi}$ is given by:

$$\frac{\partial(\log|\mathbf{\Sigma}^{-1}|)}{\partial \boldsymbol{\phi}} = 2\beta \cdot \text{diag}\left(\text{tr}(\mathbf{\Sigma T}_1), \dots, \text{tr}(\mathbf{\Sigma T}_{N_m N_d})\right)\boldsymbol{\phi} + 2\beta\left[\text{tr}(\mathbf{\Sigma U}_1), \dots, \text{tr}(\mathbf{\Sigma U}_{N_m N_d})\right]^T \tag{59}$$

Using the determinant derivative formula again, the derivatives of $\log|\mathbf{\Sigma}^{-1}|$ with respect to $\alpha_j$ and $\beta$ are given by:

$$\frac{\partial(\log|\mathbf{\Sigma}^{-1}|)}{\partial \alpha_j} = \text{tr}\left(\mathbf{\Sigma}\frac{\partial(\mathbf{A}^{-1})}{\partial \alpha_j}\right) = -\alpha_j^{-2}\mathbf{\Sigma}_{jj} \tag{60}$$

$$\frac{\partial(\log|\mathbf{\Sigma}^{-1}|)}{\partial \beta} = \text{tr}(\mathbf{\Sigma H}^T\mathbf{H}) = \text{tr}(\beta^{-1}\mathbf{\Sigma}\mathbf{\Sigma}^{-1} - \beta^{-1}\mathbf{\Sigma A}^{-1}) = \beta^{-1}\sum_{j=1}^{N_\theta}(1 - \alpha_j^{-1}\mathbf{\Sigma}_{jj}) \tag{61}$$

where $\mathbf{\Sigma}_{jj}$ is the $j^{th}$ diagonal element of the posterior covariance matrix of $\mathbf{\Sigma}$ in (20b).

Note that when differentiating (53) with respect to uncertain parameters $\boldsymbol{\phi}, \boldsymbol{\omega}^2, \beta$ and $\boldsymbol{\alpha}$, the terms involving derivatives of $\boldsymbol{\mu}$ will drop out since:

$$\beta(d\boldsymbol{\mu})^T\mathbf{H}^T(\mathbf{H}\boldsymbol{\mu} - \mathbf{b}) + \beta(\mathbf{H}\boldsymbol{\mu} - \mathbf{b})^T\mathbf{H} \cdot d\boldsymbol{\mu} + (-d\boldsymbol{\mu})^T\mathbf{A}^{-1}(\widehat{\boldsymbol{\theta}}_u - \boldsymbol{\mu}) + (\widehat{\boldsymbol{\theta}}_u - \boldsymbol{\mu})^T\mathbf{A}^{-1}\cdot(-d\boldsymbol{\mu})$$

$$= (d\boldsymbol{\mu})^T(\beta\mathbf{H}^T\mathbf{H} + \mathbf{A}^{-1})\boldsymbol{\mu} + \boldsymbol{\mu}^T(\beta\mathbf{H}^T\mathbf{H} + \mathbf{A}^{-1})d\boldsymbol{\mu} - \beta(d\boldsymbol{\mu})^T\mathbf{H}^T\mathbf{b} - \beta\mathbf{b}^T\mathbf{H}\cdot d\boldsymbol{\mu} - d\boldsymbol{\mu}^T\mathbf{A}^{-1}\widehat{\boldsymbol{\theta}}_u - \widehat{\boldsymbol{\theta}}_u^T\mathbf{A}^{-1}d\boldsymbol{\mu}$$

$$= (d\boldsymbol{\mu})^T\mathbf{\Sigma}^{-1}\boldsymbol{\mu} + \boldsymbol{\mu}^T\mathbf{\Sigma}^{-1}\cdot d\boldsymbol{\mu} - (d\boldsymbol{\mu})^T\mathbf{\Sigma}^{-1}\boldsymbol{\mu} - \boldsymbol{\mu}^T\mathbf{\Sigma}^{-1}\cdot d\boldsymbol{\mu}$$

$$= 0 \tag{62}$$

Then for the term $\beta\|\mathbf{H}\boldsymbol{\mu} - \mathbf{b}\|^2$ in (53), the derivatives with respect to $\boldsymbol{\phi}, \boldsymbol{\omega}^2$ and $\beta$ are expressed as:

$$\partial(\beta\|\mathbf{H}\boldsymbol{\mu} - \mathbf{b}\|^2)/\partial \boldsymbol{\phi} = \partial(\beta\|\mathbf{F}\boldsymbol{\phi}\|^2)/\partial \boldsymbol{\phi} = 2\beta\mathbf{F}^T\mathbf{F}\boldsymbol{\phi} \tag{63}$$

$$\partial(\beta\|\mathbf{H}\boldsymbol{\mu} - \mathbf{b}\|^2)/\partial \boldsymbol{\omega}^2 = \partial(\beta\|\mathbf{G}\boldsymbol{\omega}^2 - \mathbf{c}\|^2)/\partial \boldsymbol{\omega}^2 = 2\beta(\mathbf{G}^T\mathbf{G}\boldsymbol{\omega}^2 - \mathbf{G}^T\mathbf{c}) \tag{64}$$

$$\partial(\beta\|\mathbf{H}\boldsymbol{\mu} - \mathbf{b}\|^2)/\partial \beta = \|\mathbf{H}\boldsymbol{\mu} - \mathbf{b}\|^2 \tag{65}$$

where $\mathbf{F}$ is defined in (29); $\mathbf{G}$ and $\mathbf{c}$ are defined in (35) and (36), respectively.

The derivative of the second term $(\widehat{\boldsymbol{\theta}}_u - \boldsymbol{\mu})^T\mathbf{A}^{-1}(\widehat{\boldsymbol{\theta}}_u - \boldsymbol{\mu})$ in (53) with respect to $\boldsymbol{\alpha}$ is calculated as:

$$\partial\left((\widehat{\boldsymbol{\theta}}_u - \boldsymbol{\mu})^T\mathbf{A}^{-1}(\widehat{\boldsymbol{\theta}}_u - \boldsymbol{\mu})\right)/\partial \alpha_j = -\alpha_j^{-2}(\widehat{\boldsymbol{\theta}}_u - \boldsymbol{\mu})_j^2 \tag{66}$$

Now we differentiate the simplified $J(\boldsymbol{\delta})$ in Eq. (49) with respect to $\boldsymbol{\phi}, \boldsymbol{\omega}^2, \alpha_j$ and $\beta$ to get:

$$\frac{\partial J(\boldsymbol{\delta})}{\partial \boldsymbol{\phi}} = \frac{1}{2}\frac{\partial}{\partial \boldsymbol{\phi}}\left[-(\widehat{\boldsymbol{\psi}} - \boldsymbol{\Gamma}\boldsymbol{\phi})^T\mathbf{C}^{-1}(\widehat{\boldsymbol{\psi}} - \boldsymbol{\Gamma}\boldsymbol{\phi}) - \log|\mathbf{\Sigma}^{-1}| - \beta\|\mathbf{H}\boldsymbol{\mu} - \mathbf{b}\|^2\right]$$

$$= -(\boldsymbol{\Gamma}^T\mathbf{C}^{-1}\boldsymbol{\Gamma}\boldsymbol{\phi} - \boldsymbol{\Gamma}^T\mathbf{C}^{-1}\widehat{\boldsymbol{\psi}}) - \beta\text{diag}\left(\text{tr}(\mathbf{\Sigma T}_1), \dots, \text{tr}(\mathbf{\Sigma T}_{N_m N_d})\right)\boldsymbol{\phi} - \beta\left[\text{tr}(\mathbf{\Sigma U}_1), \dots, \text{tr}(\mathbf{\Sigma U}_{N_m N_d})\right]^T - \beta\mathbf{F}^T\mathbf{F}\boldsymbol{\phi}$$

$$\tag{67}$$

$$\frac{\partial J(\boldsymbol{\delta})}{\partial \boldsymbol{\omega}^2} = \frac{1}{2}\frac{\partial}{\partial \boldsymbol{\omega}^2}\left[-(\widehat{\boldsymbol{\omega}}^2 - \mathbf{L}\boldsymbol{\omega}^2)^T \mathbf{E}^{-1}(\widehat{\boldsymbol{\omega}}^2 - \mathbf{L}\boldsymbol{\omega}^2) - \beta\|\mathbf{H}\boldsymbol{\mu} - \mathbf{b}\|^2\right]$$

$$= -(\mathbf{L}^T\mathbf{E}^{-1}\mathbf{L}\boldsymbol{\omega}^2 - \mathbf{L}^T\mathbf{E}^{-1}\widehat{\boldsymbol{\omega}}^2) - \beta(\mathbf{G}^T\mathbf{G}\boldsymbol{\omega}^2 - \mathbf{G}^T\mathbf{c}) \tag{68}$$

$$\frac{\partial J(\boldsymbol{\delta})}{\partial \alpha_j} = \frac{1}{2}\frac{\partial}{\partial \alpha_j}\left[-(\widehat{\boldsymbol{\theta}}_u - \boldsymbol{\mu})^T \mathbf{A}^{-1}(\widehat{\boldsymbol{\theta}}_u - \boldsymbol{\mu}) - \log|\boldsymbol{\Sigma}^{-1}| - \log|\mathbf{A}|\right]$$

$$= -\frac{1}{2}\left[-\alpha_j^{-2}(\widehat{\boldsymbol{\theta}}_u - \boldsymbol{\mu})_j^2 + \alpha_j^{-1} - \alpha_j^{-2}(\boldsymbol{\Sigma_\theta})_{jj}\right] \tag{69}$$

$$\frac{\partial J(\boldsymbol{\delta})}{\partial \beta} = \frac{1}{2}\frac{\partial}{\partial \beta}\left[(N_d N_m + 2a_0 - 2)\log\beta - \log|\boldsymbol{\Sigma}^{-1}| - \beta\|\mathbf{H}\boldsymbol{\mu} - \mathbf{b}\|^2 - 2b_0^{-1}\beta\right]$$

$$= \frac{N_d N_m + 2a_0 - 2}{2}\beta^{-1} - \frac{1}{2}\|\mathbf{H}\boldsymbol{\mu} - \mathbf{b}\|^2 - \frac{1}{2}\beta^{-1}\sum_{j=1}^{N_\theta}\left(1 - \alpha_j^{-1}\boldsymbol{\Sigma}_{jj}\right) - b_0^{-1} \tag{70}$$

Setting the derivatives in (67), (68), (69) and (70) to zero leads to the update formulae for $\boldsymbol{\phi}, \boldsymbol{\omega}^2, \alpha_j$ and $\beta$ given in (26), (32), (38) and (39), respectively. Finally, by setting the derivatives of $J(\boldsymbol{\delta})$ in Eq. (49) with respect to $\eta, \nu, \boldsymbol{\rho}, \boldsymbol{\tau}, b_0$ and $\kappa$ to zero, we get their respective update formulae given in (27), (28), (33), (34), (41) and (42).

**References**

[1]. S.W. Doebling, C.R. Farrar and M.B. Prime, A summary review of vibration-based damage identification methods, Shock Vib. Digest 30(2) (1998) 91–105.

[2]. M.W. Vanik, J.L. Beck and S.K. Au, Bayesian probabilistic approach to structural health monitoring, J. Engrg. Mech. 126 (7) (2000) 738–745.

[3]. J. P. Lynch, An overview of wireless structural health monitoring for civil structures, Phil. Trans. R. Soc. A 365 (2007) 345–72.

[4]. J.P Ou and H. Li, Structural health monitoring in mainland China: review and future trends, Struct. Health Monit. 9(3) (2010) 219–231.

[5]. C.R. Farrar and K. Worden, Structural health monitoring: a machine learning perspective, John Wiley & Sons. 2013.

[6]. J.L. Beck, S.K. Au and M.W. Vanik, Monitoring structural health using a probabilistic measure, Comput.-Aided Civil Infrastruct. Eng. 16(1) (2001) 1–11.

[7]. B. Jaishia and W.X. Ren, Damage detection by finite element model updating using modal flexibility residual, Journal of Sound and Vibration, 290(1-2) (2006) 369–387

[8]. B. Moaveni, A. Stavridis, G. Lombaert, J. P. Conte, and P. B. Shing, Finite-element model updating for assessment of progressive damage in a 3-Story infilled RC frame, J. Struct. Eng. 139 (2013) 1665–1674.

[9]. D. Bernal, Load vectors for damage localization, J. Engrg. Mech. 128(1) (2002) 7–14.

[10]. J.L. Beck, Bayesian system identification based on probability logic, Struct. Control. Health Monit. 17(7) (2010) 825–847.

[11]. R. T. Cox, Probability, frequency and reasonable expectation, Am. J. Phys., 14(1) (1946) 1–13.

[12]. E.T. Jaynes, Information theory and statistical mechanics, Physical Review, 106 (1957) 620–630.

[13]. E.T. Jaynes, Probability Theory: The Logic of Science, Cambridge University Press: Cambridge, U.K., 2003.

[14]. M.W. Vanik, A Bayesian probabilistic approach to structural health monitoring, Technical Report EERL 97-07, Earthquake Engineering Research Laboratory, Caltech, Pasadena, CA, 1997.

[15]. H. Sohn and K.H. Law, A Bayesian probabilistic approach for structure damage detection, Earthquake Engrg. Struct. Dyn. 26(12) (1997) 1259–1281.

[16]. J. Ching and J.L. Beck, Two-step Bayesian structure health monitoring approach for IASC-ASCE phase II simulated and experimental benchmark studies, Technical Report EERL 2003-02, Earthquake Engineering Research Laboratory, California Institute of Technology, Pasadena, CA, 2003.

[17]. K.V. Yuen, S.K. Au and J.L. Beck, Two-stage structural health monitoring approach for phase I benchmark studies, J. Engrg. Mech. 130(2) (2004) 16–33.

[18]. J. Ching and J.L. Beck, New Bayesian model updating algorithm applied to a structural health monitoring benchmark, Struct. Health Monit. 3(4) (2004) 313–332.

[19]. K.V. Yuen, J.L. Beck and L.S. Katafygiotis, Efficient model updating and health monitoring methodology using incomplete modal data without mode matching, Struct. Control. Health Monit. 13 (1) (2006) 91–107.

[20]. H. Zhou, Y. Ni, J. Ko, Eliminating temperature effect in vibration-based structural damage detection, J.

Engrg. Mech. 137(12) (2011) 785–796.

[21]. H. Sohn, D.W. Allen, K. Worden and C.R. Farrar, Structural damage classification using extreme value statistics, J. Dyn. Sys., Meas., Control 127(1) (2005) 125–132.

[22]. M.E. Tipping, Sparse Bayesian learning and the relevance vector machine, J. Mach. Learn. Res. 1 (2001) 211–244.

[23]. D. Wipf, J. Palmer and B. Rao, Perspectives on sparse Bayesian learning, Adv. Neural Inf. Process. Syst. 16 (2004).

[24]. S. Ji, Y. Xue and L. Carin, Bayesian compressive sensing, IEEE Trans. Signal Process. 56(6) (2008) 2346–2356.

[25]. S. Babacan, R. Molina and A. Katsaggelos, Bayesian compressive sensing using Laplace priors, IEEE Trans. Image Process. 19(1) (2010) 53–64.

[26]. Y. Huang, J.L. Beck, S. Wu and H. Li, Robust Bayesian compressive sensing for signals in structural health monitoring, Comput.-Aided Civil Infrastruct. Eng. 29(3) (2014) 160–179.

[27]. D. J. C., Mackay, Bayesian methods for adaptive models. Ph.D. Thesis in Computation and Neural Systems, California Institute of Technology, Pasadena, California, 1992.

[28]. Y. Huang and J.L. Beck, Novel sparse Bayesian learning for structural health monitoring using incomplete modal data, in: Proceedings of 2013 ASCE International Workshop on Computing in Civil Engineering, Los Angeles, CA, June, 2013.

[29]. Y. Huang and J.L. Beck, Hierarchical sparse Bayesian learning for structural health monitoring with incomplete modal data, International Journal for Uncertainty Quantification, in press (2015) (DOI: 10.1615/Int.J.UncertaintyQuantification.2015011808; arXiv:1408.3685).

[30]. J.L. Beck and P.C. Jennings, Structural identification using linear models and earthquake records, Earthquake Engrg. Struct. Dyn. 8(2) (1980)145–160.

[31]. J.L. Beck and L.S. Katafygiotis, Updating models and their uncertainties. I: Bayesian statistical framework, J. Engrg. Mech. 124(4) (1998) 455–461.

[32]. S.S Chen, D.L. Donoho and M.A. Saunders, Atomic decomposition by basis pursuit, SIAM J. Sci. Stat.

Comput. 20 (1) (1999) 33–61.

[33]. E.J. Candes, J. Romberg, T. Tao, Robust uncertainty principles: Exact signal reconstruction from highly incomplete frequency information, IEEE Trans. Inf. Theory 52(2) (2006) 489–509.

[34]. J.A. Tropp and A.C. Gilbert, Signal recovery from random measurements via orthogonal matching pursuit, IEEE Trans. Inf. Theory 53(12) (2007) 4655–4666.

[35]. C. Papadimitriou, J.L. Beck and L.S. Katafygiotis, Asymptotic expansions for reliabilities and moments of uncertain dynamic systems, J. Engrg. Mech. 123 (1997) 1219–1229.

[36]. D. Bernal, S.J. Dyke, H. F. Lam and J.L. Beck, Phase II of the ASCE benchmark study on SHM, in: Proceedings of 15th Eng. Mechanics Conf., ASCE (2002) 1048–1055.

[37]. S. J. Dyke, D. Bernal, J. L. Beck and C. Ventura, Experimental Phase II of the Structural Health Monitoring Benchmark Problem, in: Proceedings of 16th Eng. Mechanics Conf., ASCE, Reston, VA, (2003).